\documentclass{aastex63}
\newcommand{\todo}{\ifmmode \text{\color{red}\Huge{\(\bullet\)}} \else {\color{red}{\Huge$\bullet$}}\fi}
\newcommand{\tido}{\ifmmode {{\color{red}\bullet}} \else {\color{red}$\bullet$}\fi}

\newcommand{\ergs}	{\ifmmode {\rm erg\,s}^{-1} \else erg s$^{-1}$\fi}
\newcommand{\kev}	{\ifmmode {\rm keV} \else keV\fi}
\newcommand{\delm}{\ifmmode \Delta \mathcal{M}_{\rm BH} \else $\Delta \mathcal{M}_{\rm BH}$\fi}

\newcommand{\CaHK}{Ca\,\textsc{ii} H+K\,$\lambda$3969, 3934 }
\newcommand{\MgI}{Mg\,\textsc{i}\,$b\,\lambda$5183, 5172, 5167 }

\newcommand{\cahk}{Ca\,H+K }
\newcommand{\mgi}{Mg\,\textsc{i}}
\newcommand{\cat}{CaT }
\newcommand{\sigs}{\ifmmode \sigma_\mathrm{\star} \else $\sigma_\mathrm{\star}$\fi}
\newcommand{\sig}{\sigs}
\newcommand{\sigr}{\ifmmode \sigma_{\rm Red} \else $\sigma_{\rm Red}$\fi}
\newcommand{\sigb}{\ifmmode \sigma_{\rm Blue} \else $\sigma_{\rm Blue}$\fi}
\newcommand{  \Lbol     }{\ifmmode L_{\rm bol} \else $L_{\rm bol}$\fi}
\newcommand{\LHa}{\ensuremath{L_{\rm H\alpha}}} 
\newcommand{\LHaobs}{\ensuremath{L^{\rm obs}_{\rm H\alpha}}}  
\newcommand{\LHacorr}{\ensuremath{L^{\rm corr}_{\rm H\alpha}}}  
\newcommand{\av}{\ifmmode A_{\rm V} \else $A_{\rm V}$\fi}
\newcommand{\AV}{\av}
\newcommand{\AHa}{\ifmmode A_{\rm H\alpha} \else $A_{\rm H\alpha}$\fi}
\newcommand {\Lhard} {\ensuremath{L_{\rm 2-10\,\kev}}}

\newcommand {\Lhardint} {\ensuremath{L^{\rm int}_{\rm 2-10\,\kev}}}
\newcommand{\cc}	{\ifmmode {\rm cm}^{-3}    \else cm$^{-3}$\fi}
\newcommand{\cmii}	{\ifmmode {\rm cm}^{-2}    \else cm$^{-2}$\fi}

\newcommand {\eddr} {\ifmmode \log \lambda_\mathrm{Edd} \else $\log \lambda_\mathrm{Edd}$\fi}
\newcommand {\eddrc}{\ifmmode \log \lambda_\mathrm{Edd, corr} \else $\log \lambda_\mathrm{Edd, corr}$\fi}
\newcommand{\Halpha}{\ifmmode {\rm H}\alpha \else H$\alpha$\fi}
\newcommand{\halpha}{\Halpha}
\newcommand{\ha}{\Halpha}
\newcommand{\Hbeta}{\ifmmode {\rm H}\beta \else H$\beta$\fi}

\newcommand{\hb}{\Hbeta}
\newcommand{\oiii}{\ifmmode \left[{\rm O}\,\textsc{iii}\right] \else [O\,{\sc iii}]\fi}
\newcommand{\OIII}{\ifmmode \left[{\rm O}\,\textsc{iii}\right]\,\lambda5007 \else [O\,{\sc iii}]\,$\lambda5007$\fi}
\newcommand{\nii}{\ifmmode \left[{\rm N}\,\textsc{ii}\right]  \else [N\,\textsc{ii}]\fi}
\newcommand{\NII}{\ifmmode \left[{\rm N}\,\textsc{ii}\right]\,\lambda6584 \else [N\,\textsc{ii}]\,$\lambda6584$\fi}

\newcommand {\nh}{\ifmmode N_{\rm H} \else $N_{\rm H}$\fi}

\newcommand {\Lsoftint} {\ifmmode L^{\rm in}_{\mathrm{2-10\ keV}} \else $L^{\rm in}_{\mathrm{2-10\ keV}}$\fi}

\newcommand {\ergpersec} {\ifmmode {\rm erg~s}^{-1} \else erg~s$^{-1}$ \fi}

\def\arcsec{{\mbox{$^{\prime \prime}$}}}

\def\arcsec{{\mbox{$^{\prime \prime}$}}}

\def\g{{\rm\thinspace g}}

\def\K{{\rm\thinspace K}}

\def\km{{\rm\thinspace km}}

\def\m{{\rm\thinspace m}}
\def\Mpc{{\rm\thinspace Mpc}}
\newcommand{\Msun}{\ifmmode \rm M_{ \odot} \else \rm M$_{ \odot}$\fi}

\def\s{{\rm\thinspace s}}

\def\kmps{\hbox{$\km\s^{-1}\,$}}

\def\kmpspMpc{\hbox{$\kmps\Mpc^{-1}$}}

\def\apjl{\hbox{ApJL}}
\def\apjs{\hbox{ApJS}}

\def\aap{\hbox{AAP}}

\def\arcsec{{\mbox{$^{\prime \prime}$}}}

\newcommand{\lledd}{\ifmmode L/L_{\rm Edd} \else $L/L_{\rm Edd}$\fi}
\newcommand{\mbh}{\ifmmode M_{\rm BH} \else $M_{\rm BH}$\fi}
\newcommand{\logmbh}{\ifmmode \log M_{\rm BH / M\odot} \else $\log M_{\rm BH / M\odot}$\fi}
\newcommand{\kms}{\ifmmode {\rm km\,s}^{-1} \else ${\rm km\,s}^{-1}$\fi}
\newcommand{\nuvr}{\ifmmode {\rm NUV}-r \else NUV-$r$\fi}
\newcommand{\mh}{\ifmmode M_{\rm H_2} \else $M_{\rm H_2}$\fi}
\newcommand{\mhi}{\ifmmode M_{\rm HI} \else $M_{\rm HI}$\fi}
\newcommand{\mstar}{\ifmmode M_{\ast} \else $M_{\ast}$\fi} 
\newcommand{\must}{\ifmmode \mu_{\ast} \else $\mu_{\ast}$\fi}
\newcommand{\hmol}{\ifmmode H_2 \else $H_2$\fi}
\newcommand{\rmol}{\ifmmode R_{\rm mol} \else $R_{\rm mol}$\fi}
\newcommand{\tdep}{\ifmmode t_{\rm dep}({\rm H_2}) \else $t_{\rm dep}({\rm H_2})$\fi}
\newcommand{\tdepHI}{\ifmmode t_{\rm dep}({\rm HI}) \else $t_{\rm dep}({\rm HI})$\fi}
\newcommand{\fgas}{\ifmmode f_{\rm H_2} \else $f_{\rm H_2}$\fi}
\newcommand{\fhi}{\ifmmode f_{\rm HI} \else $f_{\rm HI}$\fi}
\newcommand{\xco}{\ifmmode \alpha_{\rm CO} \else $\alpha_{\rm CO}$\fi}

\usepackage{tablefootnote}
\usepackage{python}
\usepackage{graphicx}
\usepackage{xcolor}
\usepackage{lineno}
\usepackage{booktabs}
\accepted{by the ApJ}

\shorttitle{The $\mbh - \sig$ Relation of Type 1 AGNs}
\shortauthors{Caglar et al.}

\begin{document}

\title{BASS XXXV. The $\mbh - \sig$ Relation of 105-Month Swift-BAT Type 1 AGNs}  

\suppressAffiliations
\correspondingauthor{Turgay Caglar}
\email{caglar@strw.leidenuniv.nl}

\author[0000-0002-9144-2255]{Turgay Caglar}
\affiliation{Leiden Observatory, PO Box 9513, 2300 RA, Leiden, the Netherlands}

\author[0000-0002-7998-9581]{Michael J. Koss}
\affiliation{Eureka Scientific, 2452 Delmer Street Suite 100, Oakland, CA 94602-3017, USA}
\affiliation{Space Science Institute, 4750 Walnut Street, Suite 205, Boulder, CO80301, USA}

\author[0000-0003-1014-043X]{Leonard Burtscher}
\affiliation{Leiden Observatory, PO Box 9513, 2300 RA, Leiden, the Netherlands}

\author[0000-0002-3683-7297]{Benny Trakhtenbrot}
\affiliation{School of Physics and Astronomy, Tel Aviv University, Tel Aviv 69978, Israel}

\author[0000-0002-6770-5043]{M. Kiyami Erdim}
\affiliation{Yildiz Technical University, Graduate School of Natural and Applied Sciences, Istanbul 34220, Turkey}

\author[0000-0001-8450-7463]{Julian E. Mej\'ia-Restrepo}
\affiliation{European Southern Observatory, Casilla 19001, Santiago 19, Chile}

\author[0000-0001-5742-5980]{Federica Ricci}
\affiliation{Dipartimento di Matematica e Fisica, Università degli Studi Roma Tre, via della `
Vasca Navale 84, 00146, Roma, Italy}
\affiliation{INAF - Osservatorio Astronomico di Roma, via Frascati 33, 00044 Monte Porzio Catone, Italy}

\author[0000-0003-2284-8603]{Meredith C. Powell}
\affiliation{Kavli Institute for Particle Astrophysics and Cosmology, Stanford University, 452 Lomita Mall, Stanford, CA 94305}

\author[0000-0001-5231-2645]{Claudio Ricci}
\affiliation{N\'ucleo de Astronom\'ia de la Facultad de Ingenier\'ia, Universidad Diego Portales, Av. Ej\'ercito Libertador 441, Santiago, Chile}
\affiliation{Kavli Institute for Astronomy and Astrophysics, Peking University, Beijing 100871, People's Republic of China}
\affiliation{George Mason University, Department of Physics \& Astronomy, MS 3F3, 4400 University Drive, Fairfax, VA 22030, USA}

\author[0000-0002-7962-5446]{Richard Mushotzky}
\affiliation{Department of Astronomy and Joint Space-Science Institute, University of Maryland, College Park, MD 20742, USA}

\author[0000-0002-8686-8737]{Franz E. Bauer}
\affiliation{Instituto de Astrof\'{\i}sica  and Centro de Astroingenier{\'{\i}}a, Facultad de F\'{i}sica, Pontificia Universidad Cat\'{o}lica de Chile, Casilla 306, Santiago 22, Chile}
\affiliation{Millennium Institute of Astrophysics (MAS), Nuncio Monse{\~{n}}or S{\'{o}}tero Sanz 100, Providencia, Santiago, Chile}
\affiliation{Space Science Institute, 4750 Walnut Street, Suite 205, Boulder, CO 80301, USA}

\author[0000-0001-8211-3807]{Tonima T. Ananna}
\affiliation{ Department of Physics and Astronomy, Dartmouth College, 6127 Wilder Laboratory, Hanover, NH 03755, USA}

\author[0000-0001-5481-8607]{Rudolf E. B\"ar}
\affiliation{Institute for Particle Physics and Astrophysics, ETH Z\"urich, Wolfgang-Pauli-Strasse 27, CH-8093 Z\"urich, Switzerland}

\author[0000-0001-9737-169X]{Bernhard Brandl}
\affiliation{Leiden Observatory, PO Box 9513, 2300 RA, Leiden, the Netherlands}

\author[0000-0003-4359-8797]{Jarle Brinchmann}
\affiliation{Centro de Astrof\'isica da Universidade do Porto, Rua das Estrelas, 4150-762 Porto, Portugal}
\affiliation{Instituto de Astrof\'isica e Ci\^{e}ncias do Espaço, Universidade do Porto, CAUP, Rua das Estrelas, 4150-762 Porto, Portugal}

\author{Fiona Harrison}
\affiliation{Cahill Center for Astronomy and Astrophysics, California Institute of Technology, Pasadena, CA 91125, USA}

\author[0000-0002-4377-903X]{Kohei Ichikawa}
\affiliation{Astronomical Institute, Tohoku University, Aramaki, Aoba-ku, Sendai, Miyagi 980-8578, Japan}
\affiliation{Frontier Research Institute for Interdisciplinary Sciences, Tohoku University, Sendai 980-8578, Japan}

\author[0000-0002-2603-2639]{Darshan Kakkad}
\affiliation{Space Telescope Science Institute, 3700 San Martin Drive, Baltimore, MD 21218, USA}

\author[0000-0002-5037-951X]{Kyuseok Oh}
\affiliation{Korea Astronomy \& Space Science institute, 776, Daedeokdae-ro, Yuseong-gu, Daejeon 34055, Republic of Korea}
\affiliation{Department of Astronomy, Kyoto University, Kitashirakawa-Oiwake-cho, Sakyo-ku, Kyoto 606-8502, Japan}

\author[0000-0002-1321-1320]{Rog\'erio Riffel}
\affiliation{Departamento de Astronomia, Universidade Federal do Rio Grande do Sul. Av. Bento Gonçalves 9500, 91501-970 Porto Alegre, RS, Brazil}
\affiliation{Instituto de Astrof\'\i sica de Canarias, Calle V\'\i a L\'actea s/n, E-38205 La Laguna, Tenerife, Spain}

\author[0000-0001-8020-3884]{Lia F. Sartori}
\affiliation{Institute for Particle Physics and Astrophysics, ETH Z\"urich, Wolfgang-Pauli-Strasse 27, CH-8093 Z\"urich, Switzerland}

\author[0000-0001-5785-7038]{Krista L. Smith}
\affiliation{Department of Physics and Astronomy, Texas A\&M University, College Station, TX 77843, USA}

\author[0000-0003-2686-9241]{Daniel Stern}
\affiliation{Jet Propulsion Laboratory, California Institute of Technology, 4800 Oak Grove Drive, MS 169-224, Pasadena, CA 91109, USA}

\author[0000-0002-0745-9792]{C. Megan Urry}
\affiliation{Yale Center for Astronomy \& Astrophysics and Department of Physics, Yale University, P.O. Box 208120, New Haven, CT 06520-8120, USA}

\begin{abstract}

We present two independent measurements of stellar velocity dispersions ( \sig\ ) from the \CaHK \& \MgI region (3880--5550~\AA) and the Calcium Triplet region (CaT, 8350--8750~\AA) for 173 hard X-ray-selected Type 1 AGNs ($z \leq$ 0.08) from the 105-month Swift-BAT catalog. We construct one of the largest samples of local Type 1 AGNs that have both single-epoch (SE) 'virial' black hole mass (\mbh) estimates and \sigs\ measurements obtained from high spectral resolution data, allowing us to test the usage of such methods for SMBH studies. We find that the two independent \sigs\ measurements are highly consistent with each other, with an average offset of only $0.002\pm0.001$ dex. Comparing \mbh\ estimates based on broad emission lines and stellar velocity dispersion measurements, we find that the former is systematically lower by $\approx$0.12 dex. Consequently, Eddington ratios estimated through broad-line \mbh\ determinations are similarly biased (but in the opposite way). We argue that the discrepancy is driven by extinction in the broad-line region (BLR). We also find an anti-correlation between the offset from the $\mbh - \sig$ relation and the Eddington ratio. Our sample of Type 1 AGNs shows a shallower $\mbh - \sig$ relation (with a power law exponent of $\approx$3.5) compared with that of inactive galaxies (with a power-law exponent of $\approx$4.5), confirming earlier results obtained from smaller samples. 
\end{abstract}


\section{Introduction}\label{sec_intro}

Supermassive black holes (SMBHs), residing in the centers of massive galaxies, are commonly thought to co-evolve with their host galaxies, as demonstrated by the present-day correlations between SMBH mass (\mbh) and several host properties \citep[e.g.,][]{MM13, koho13, Saglia}, such as the stellar velocity dispersion ( \sig\ ) \citep[][]{Ferrarese,Geb,Merri,Tremaine,Gultekin}, bulge luminosity \citep[][]{Kormendy95,Marconi} and bulge mass \citep[][]{Magorrian,Haring}. Additionally, correlations with the bulge average spherical density, half-mass radius \citep[e.g.,][]{Saglia} and dark matter halos \citep[e.g.,][]{2002Ferrarese,2003Baes,2009Bandara,2011Volonteri,2015Sabra,2021Marasco,2022Merry}, have been proposed. Of these, the $\mbh - \sig$ relation is still the tightest relation among them \cite[with an intrinsic scatter of $\sim$0.3 dex; e.g.,][]{Gultekin,Saglia,VandenB}.

The observed close relations between \mbh\ and host properties strongly support a co-evolutionary scenario, where some form of `feedback' exerted by actively accreting SMBHs (i.e., active galactic nuclei; hereafter AGNs) affects the host galaxy growth. Indeed, specific feedback models have been shown to be related to the shape of the $\mbh - \sig$ relation, and in particular its exponent, $\beta$ (where $\mbh \propto\sig^\beta$), with $\beta \simeq 4$ attributed for momentum-driven feedback and $\beta \simeq 5$ attributed for energy-driven feedback \citep[][respectively]{Silk,King}. 
Observationally, the details and impact of AGN feedback, as well as the slope $\beta$ are not yet settled. \citet[][hereafter KH13]{koho13} find  $\beta=4.38$ using spatially resolved gas and stellar kinematics for elliptical and classical bulge hosting local galaxies. However, \citet[][henceforth MM13]{MM13} report an $\beta=5.64$ using $\mbh$ estimates from spatially resolved dynamics in a sample of local early- and late-type galaxies (including brightest cluster galaxies). Both these samples and indeed most samples used for such studies, are dominated by inactive galaxies with the presence of just a few low-luminosity AGNs. Finally, \citet{VandenB} report an $\beta = 5.35$ using a sample of galaxies in which SMBH masses are compiled using four different methods: gas dynamics, stellar dynamics, reverberation mapping, and mega-masers. 

Recent studies have suggested whether SMBH-host relations may depend on a variety of factors, including host morphology. Specifically, early- vs. late-type galaxies \citep[e.g.,][]{Gultekin,MM13}, pseudo- vs. real bulges \citep[e.g.,][]{Greene10,koho13,Ho14}, and barred vs. unbarred galaxies \citep[e.g.,][]{2008Hu,Graham2008,Graham2009,2014Hartmann} have all been discussed as possible influencing factors. Additionally, \citet{2011Xiao} found a small offset caused by the disk inclination. Interestingly, some studies have shown a $\beta \approx 3$ for galaxies with pseudo-bulges \citep[e.g.,][]{Greene10,Ho14}. However, pseudo-bulge hosting galaxies are found to have an order of magnitude lower black hole masses relative to elliptical bulge hosting galaxies \citep[e.g.,][]{Greene10,Ho14}, thus reside significantly below the $\mbh - \sig$ relation of inactive galaxies, and they also show a larger scatter at the lower-mass end. 

For actively accreting SMBHs (i.e., AGNs) the primary approach for \mbh determination is the reverberation mapping (RM) of broad emission lines \citep[e.g.,][]{Blandford,Peterson93,Onken02,Denney06,Denney10,Bentz06,Bentz09a,Bentz09b,Bentz16,Villafa}. Despite several dedicated campaigns during the past few decades, the number of reliable \mbh\ determination remains limited to $\approx$90 systems \citep[see the RM black hole mass archive;][]{Bentzarchive}. Several recent and ongoing RM campaigns aim to significantly increase the number of RM-based \mbh\ measurements, such as OzDES-RM \citep{OzdesRM}, SDSS-RM \citep{SDSSRM,SDSSRM2}, and SDSS-V \citep{2017Kollmeier}.

The so-called single-epoch (SE) $\mbh$ estimation method provides a potential solution to estimate \mbh\ for the much larger spectroscopic data sets of Type 1 AGNs, including luminous quasars that can be traced to $z\gtrsim7$. The method uses the width of the broad emission lines (either full width at half maximum; hereafter $FWHM$ or standard deviation $\sigma$) as a proxy for the virialized broad-line-region (BLR) gas velocities and the AGN continuum luminosity as a probe of the broad-line region radius. The latter is based on relatively tight correlations between BLR size and AGN continuum luminosity in various spectral bands \cite[e.g.,][]{Koratkar,Kaspi2000,Kaspi2005,Bentz06a,Bentz09a,Zaja2020}. 
Since the BLR geometry and (detailed) radiative physics are not entirely known (per source), the SE method adopts an order-of-unity scaling factor to yield \mbh. Crucially, this scaling factor, called the `virial factor' ($f$) is typically derived by \emph{assuming} that AGNs follow the same $\mbh - \sig$ relation as the one determined for inactive galaxies. Indeed, the systematic uncertainty of \mbh\ estimates derived using the SE method can reach $\gtrsim$ 0.4 dex \citep[][]{Pancoast2014,2017RicciF,Caglar}, mostly due to the intrinsic scatter in the $\mbh - \sig$ relation. 

An average virial factor of $f_\mathrm{FWHM}\approx1$ is reported with an uncertainty of 0.15 dex by calibrating RM-based $\mbh$ estimations to various versions of the $\mbh - \sig$ relation \citep[e.g.,][]{Onken04,2012ParkPark,Grier2013,Woo13,Woo15,2017Grier}. Several studies have investigated in detail the $\mbh - \sig$ relation for AGNs by combining RM-based virial \mbh\ determinations and host \sig\ measurements \citep[e.g.,][]{Nelson04,Onken04, Woo10, Woo13, Woo15, Graham11,2012ParkPark, Batiste17,Bennert}. Generally, the slope of the $\mbh - \sig$ relation for RM AGNs is found to be shallower than that of inactive galaxies \citep[$\beta \lesssim 4$; e.g.,][]{Woo13,Woo15,Bennert}, but the discrepancy between the two relations is often attributed to unreliable \sig\ measurements in AGN-dominated spectra of Type 1 AGNs, as well as due to sample selection bias \cite[e.g.,][]{Greene2006,2007Lauer,2013Shen,Shankar2016}. Interestingly, C20 has proposed that some part of the discrepancy might be caused by extinction in the BLR, which was also claimed by the following studies \citep{Ricci_DR2_NIR_Mbh,Mejia_Broadlines}. Hence, the discrepancies between the $\mbh - \sig$ relations of active and inactive galaxies may reflect real, though yet unclear, astrophysical differences between these populations. Clearly, detailed analyses of large and highly complete AGN samples and inactive galaxies are needed to address the origin of such discrepancies. 

In this work, we present stellar velocity dispersion measurements for a sample of broad-line, Type 1 AGNs drawn from the second data release of the Swift/BAT AGN Spectroscopic Survey \citep[BASS DR2\footnote{www.bass-survey.com} ,][]{Koss_DR2_overview}. 
BASS is a highly complete survey of ultra-hard X-ray-selected AGNs, mainly in the local universe. The ultra-hard X-ray selection (14--195\, keV) allows us to overcome biases related to (circumnuclear) obscuration \citep[e.g.,][]{Ricci15,Ricci17b}, host properties, etc., thus potentially circumventing some of the challenges faced by previous studies. 
We aim to study the $\mbh - \sig$ relation for our sample and compare our results with the $\mbh - \sig$ relation for inactive galaxies. We additionally investigate a potential discrepancy between two black hole mass estimates obtained from the single-epoch method versus the ones from the $\mbh - \sig$ relation. Throughout this paper, we define this discrepancy as the offset from the $\mbh - \sig$ relation as follows: $\delm \equiv\log{M_\mathrm{BH,BLR}}$ - $\log{M_\mathrm{BH,\sigma}}$. Finally, we aim to understand how such discrepancies may depend on several key AGN properties. 
This paper is organized as follows. In Section \ref{sec_sample} we introduce the BASS-based AGN sample and archival data. In Section \ref{sec:analysis} we describe our analysis methodology, while in Section \ref{sec:results} we present and discuss our main results. We conclude with a summary of our key findings in Section \ref{sec:summary}. Throughout this paper, we assume a standard flat $\Lambda$CDM cosmology, with $H_\mathrm{0}$ = 70 $\kmpspMpc$, $\Omega_\mathrm{M} = 0.3$ and $\Omega_\mathrm{\Lambda} = 0.7$.

\section{BASS Sample and Archival Data}\label{sec_sample}

The 70-month data release of Swift-BAT hard X-ray (14 -- 195\,\kev) all-sky survey \citep{Baumgartner} consists of 858 AGNs. 
The BAT AGN Spectroscopic Survey (BASS) aims to obtain optical spectroscopy for BAT-selected AGNs. 
Specifically, BASS DR2 includes optical spectra for essentially all 70-month BAT catalog AGNs, except for six highly extincted sources located at Galactic latitudes $|b|< 10\deg$. 
We also use BASS-led spectroscopy of AGNs drawn from the 105-month BAT survey \citep{Oh18}. Although this effort is not yet complete and does not represent a flux-limited sample, the spectra in hand allow us to probe fainter sources, extending the range in SMBH mass and/or Eddington ratio under study.

The targeted optical spectroscopy pursued by BASS typically covers a wide spectral range (3000-10000~\AA), in order to study both AGN-dominated broad \& narrow emission lines \citep[e.g.,][]{Mejia_Broadlines,2022Oh,Ricci_DR2_NIR_Mbh} and host galaxy properties \citep[e.g.,][]{Koss_DR2_sigs,2022Merry}. 
Key technical aspects of the spectra used for our work are provided in Section~\ref{sec:sigs_est}, where we detail our spectral measurements.
We stress that the ultra-hard X-ray Swift-BAT survey allows us to detect AGNs with a wide range of neutral hydrogen absorbing columns, ranging from unabsorbed ($\log(\nh/\cmii) = 20.0$) to Compton-thick ($\log(\nh/\cmii) > 24.0$) sources. 
Indeed, the BASS sample was shown to be less biased compared to other surveys with respect to obscuration \cite[e.g.,][]{2016Kossbass,Ricci15,Ricci22}, star formation \cite[e.g.,][]{Ichikawa17,Ichikawa19,Shimizu15}, and host molecular gas content \citep{KossMolecular}. More detailed information about BASS DR2 can be found in the main BASS DR2 overview and catalog papers \citep{Koss_DR2_overview,Koss_DR2_catalog}.

\subsection{Our Sample}

The BASS DR2 sample comprises 858 AGNs: 359 of which are classified as Type 1 sources, 393 Type 2s (including Seyfert 1.9 sources), and 106 beamed and/or lensed AGNs \citep[see][for more details]{Koss_DR2_overview,Koss_DR2_catalog}. Importantly for this work, we note that the velocity dispersion measurements for the obscured AGNs in BASS DR2 (Seyfert 1.9 and 2 AGNs) are presented in \cite{Koss_DR2_sigs}. Here we focus only on Type 1 AGNs with redshifts $z \leq 0.08$, where the redshift threshold is chosen in order to avoid telluric absorption across the CaT absorption complex. Furthermore, we excluded 40 Type 1 AGN spectra observed with low-resolution spectral setups ($R < 1000$), where \sigs\ measurements would be unreliable. Our final sample thus consists of a total of 240 AGNs, of which 185 are from the 70-month BAT catalog and 55 are a 'bonus' sample from the 105-month catalog. 

\subsection{BASS Archival Data}

\subsubsection{X-ray Data}\label{sec_extra_bass_data}

We adopted hydrogen column density measurements ($\nh$) and intrinsic (absorption-corrected) X-ray luminosity measurements, and related uncertainties (90\% confidence intervals) directly from \citet{Ricci17a} for the 70-month Swift-BAT sources in our sample. These are obtained by fitting the X-ray spectra with a variety of models, including an absorbed cutoff power-law component, an unobscured reflection component, and another cutoff power-law component for scattering. We note that there are no $\nh$ or intrinsic X-ray luminosity determinations available for the bonus sample of AGNs from the 105-month Swift-BAT catalog.

\subsubsection{Black Hole Masses}

We adopt broad-line-based SE (`virial') black hole mass estimates for our sample of AGNs from the respective BASS DR2 catalog of \cite{Mejia_Broadlines}. 
That study performed a detailed spectral decomposition and emission line fitting procedure, following \cite{Trakhtenbrot2012} and \cite{MejiaRestrepo2016a}. 
In \citet{Mejia_Broadlines}, \mbh\ is calculated  using the prescriptions calibrated by \citet[][for H$\alpha$]{GreeneHo2005} and \citet[][for H$\beta$]{Trakhtenbrot2012}, and using the $FWHM$ of the emission lines and a virial factor of $f = 1$. 
The latter choice results in a somewhat revised \mbh\ prescription (i.e. compared with the one presented in \citealt{GreeneHo2005}), of the form:
\begin{equation}\label{eq:mbh_ha}
\mbh  = 2.67 \times 10^{6} \times \left( \frac{\LHa}{10^{42}\,\ergpersec} \right)^{0.55} \times \left(\frac{{FWHM_{\mathrm H\alpha}}}{10^{3}\,\kms} \right)^{2.06} \: \Msun \, , 
\end{equation}
where the H$\alpha$ related quantities reflect only the broad emission component.
A detailed explanation of the fitting procedure and this \mbh\ prescription, as well as a complete catalog of the best-fitting parameters, can be found in \citet{Mejia_Broadlines}. Here, we note that \citet{Mejia_Broadlines} only corrected their \mbh\ estimates for the Galactic extinction, but in this work, we will even further correct their \mbh\/ estimates for the BLR extinction, which will be described in Section \ref{sec_extinction}.

\section{Analysis and Methods}\label{sec:analysis}

\subsection{Stellar Velocity Dispersion Measurements}\label{sec:sigs_est}

We measure host galaxy stellar velocity dispersion for our sample of Type 1 AGNs using the penalized pixel-fitting procedure \cite[pPXF;][]{Cappellari2004,Cappellari2017}. The pPXF routine applies the Gauss-Hermite parameterization for the line-of-sight velocity distribution in pixel space. By using pPXF, the continuum can be matched using additive polynomials, whereas bad pixels and/or emission lines can be masked from the spectra. Finally, pPXF makes initial guesses for \sigs\ by broadening the stellar templates. During this procedure, several parameters are being fit simultaneously, including the systemic velocity ($V$), the velocity dispersion ($\sigma$), and a series of Hermite polynomials, $h_\mathrm{3} \dots h_\mathrm{m}$. 

In our study, we performed the pPXF method allowing the following parameters as free: the systemic velocity ($V$), the velocity dispersion ($\sigma$), and two Hermite polynomials ($h_\mathrm{3}$ and $h_\mathrm{4}$). We supplemented pPXF with a grid of stellar spectral templates based on VLT/X-shooter observations \citep{ChenXshoo,Gonneau2020} using the velocity scale ratio of 2 that corresponds to the templates at twice the resolution of the observed spectra. The X-shooter Spectral Library we used contains 830 stellar spectra of 683 stars covering the wavelength range 3500 -- 24800~\AA\ with an average instrumental resolution of 0.51~\AA\ for the bluer spectral regions of interest (3800--5500~\AA) and 0.78~\AA\ for the red spectral region (8300--8800~\AA). However, our sample of galaxies is observed by various instruments with a variety of spectral resolutions ranging from 2.0~\AA\ to 6.0 ~\AA\ (3800-5500~\AA)\footnote{We also have spectral data from low-resolution observation setups ($R < 1000$), which are not taken into consideration for \sigs\ measurements \citep[see][]{Koss_DR2_catalog}.}.

Therefore, the template spectra were convolved with a relative line-spread function. We masked several prominent, mostly AGN-dominated emission lines that are present in our spectral regions of interest (H$\beta$,$\gamma$, $\delta$,$\epsilon$, [Ne\,\textsc{iii}]\,$\lambda$ 3968, [O\,\textsc{iii}\,$\lambda\lambda$4959,5007], O\,\textsc{i} $\lambda$8446, and Fe\,\textsc{ii} $\lambda$8616), as well as bad pixels (if these exist), to increase the robustness of our \sigs\ measurements. To mask broad emission components, we additionally applied a mask function with a range of width 2000--3500\,\kms, which is appropriate for the BLR-related widths of our broad-line AGNs. We performed pPXF fitting adopting additive (between degree = 2-8) and multiplicative (mdegree = 0-1) polynomials to develop the best match between the composite stellar population and the galaxy spectrum. We finally selected the best-fit result with the least possible degree of polynomials. To estimate the uncertainties associated with \sigs\ measurements, we used suggested residuals bootstrapping procedure \citep{2022Cappellari}. Briefly, for each AGN, we re-sample the residuals of initial \sigs\ fit to generate 100 mock spectra to perform 100 additional fits, resulting in a distribution of \sigs\ measurements as well as uncertainties in the distribution of the weights. The same bootstrapping approach was also used by \citet{Koss_DR2_sigs} for DR2 type 2 (Seyfert 1.9s and 2s) AGNs.

In Table \ref{table_mainresult}, we present the resulting \sigs\ measurements from the spectral region covering Ca\,H+K and Mg\,\textsc{i}, and/or the CaT features (henceforth \sigb\ and \sigr\, respectively) for our sample of AGNs. Additionally, the pPXF model fits are shown in Appendix \ref{sec:sigb_sigr_plots} (Figures~\ref{fig:sigb_sigr_plots}--\ref{fig:sigb_sigr_plots_fail}). 
Three experienced co-authors (T.C., L.B., and M.K.) have visually inspected the spectral fits and assigned a quality flag for each spectral region of each AGN: 0 for good quality fits, 1 for acceptable fits, 2 for unaccepted fits, and 9 for failed fits (see Table \ref{table_mainresult_fail}. For most sources, we were able to fit both the blue part and the red part of the spectra to obtain independent \sigs\  measurements. For the SOAR spectra, we only fit the \cat features region since the instrumental setup only covers that spectral region. Here, we note that, for some of our AGNs, we have multiple spectra obtained with different instruments. In Appendix \ref{sec:sigb_sigr_plots_other_instruments}, we show a comparison of \sigs\ measurements from the different instruments.

\begin{figure*}[!ht]
\centering
\includegraphics[width=8.9cm, height=6.2cm]{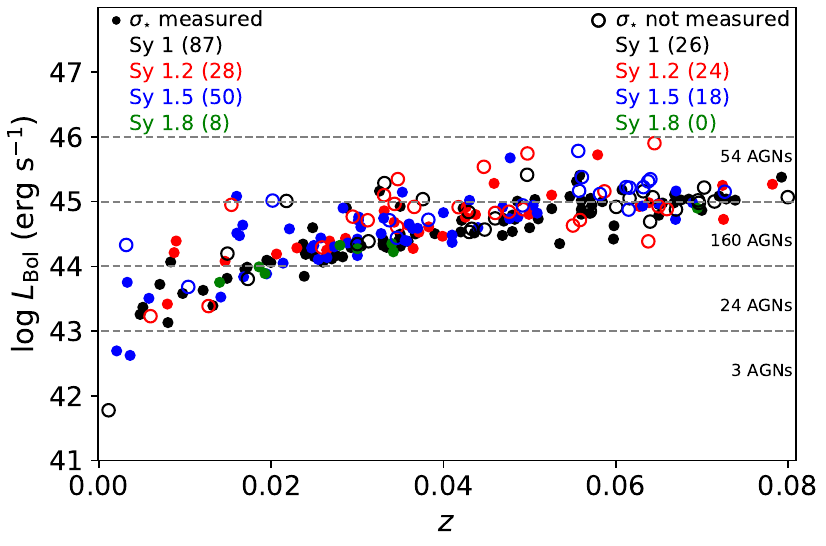}
 \includegraphics[width=8.9cm, height=6.2cm]{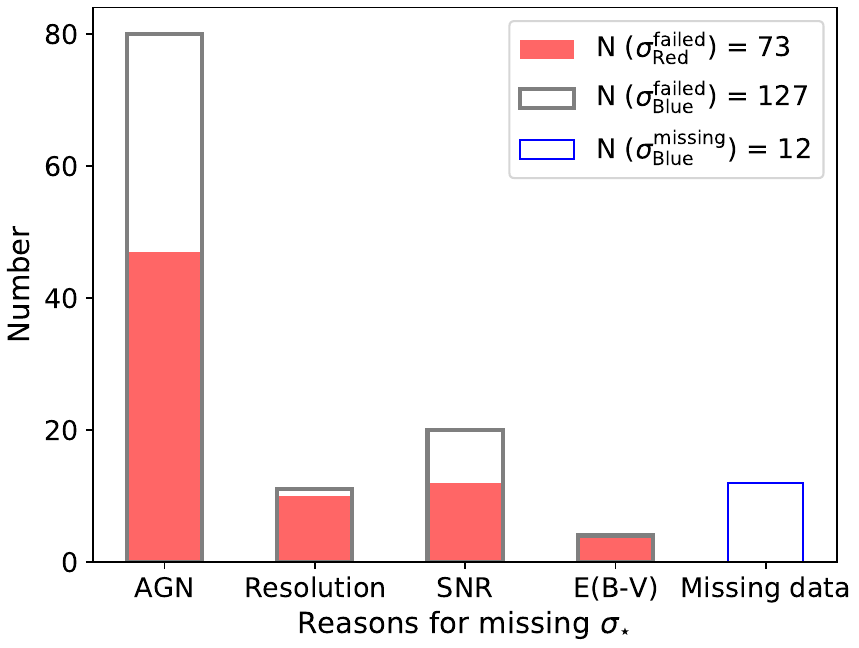}
\includegraphics[width=8.9cm, height=6.2cm]{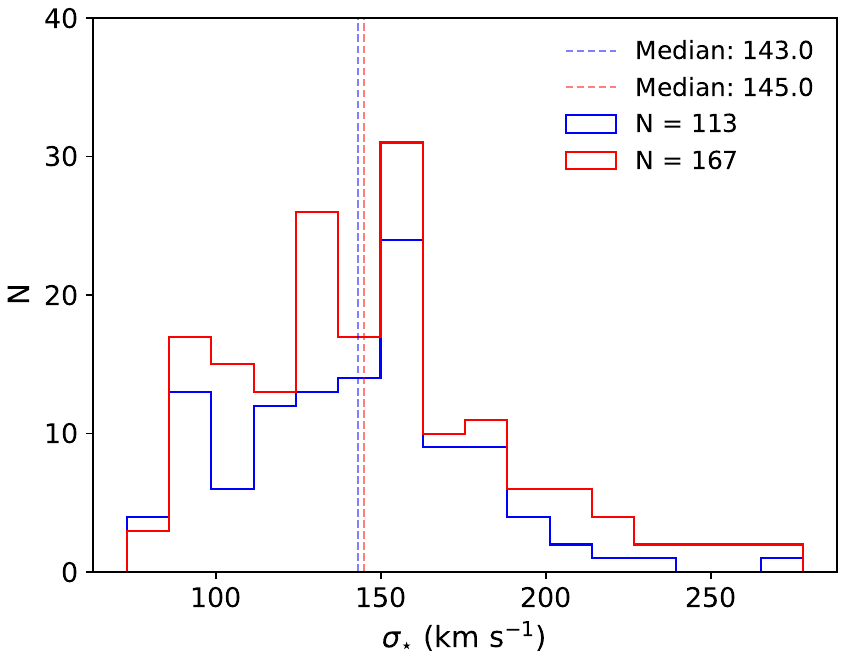}
\includegraphics[width=8.9cm, height=6.2cm]{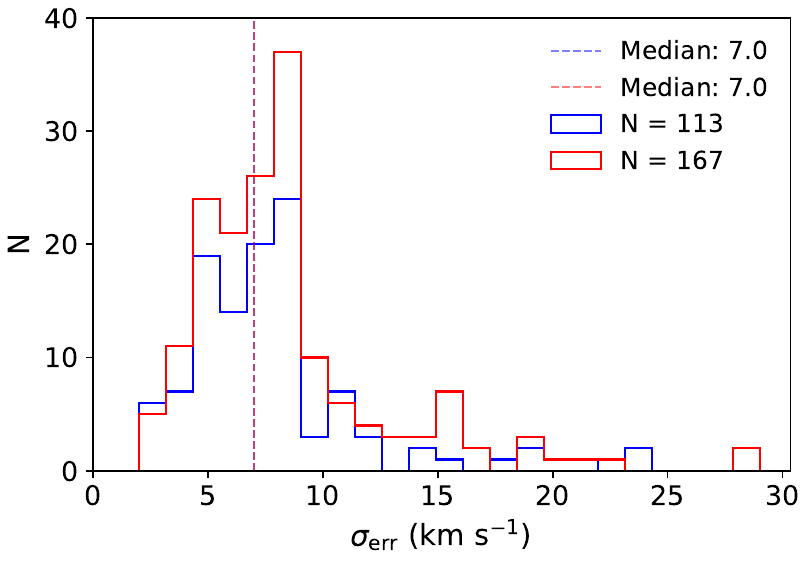}
\caption{\textbf{Top left:} the bolometric AGN luminosities versus redshift for our sample of AGNs. We show AGNs with successful \sig\ fits with filled circles, while those with failed \sig\ fits are shown with open circles. \textbf{Top right:} distributions of failed \sigb\ (based on the region covering the Ca H+K and Mg I absorption lines) and \sigr\ (based on the region that covers the Calcium triplet absorption lines) measurements are separated by failure reason. \textbf{Bottom left:} distributions of successful velocity dispersion fit results. \textbf{Bottom right:} distributions of measurement errors for successful velocity dispersion measurements. Blue histograms represent $\sigma_\mathrm{Blue}$, whereas red histograms represent \sigr\/. The vertical blue and red lines correspond to the median values for \sigb\ and \sigr\/, respectively. 
}
\label{fig_sig_hist}
\end{figure*}

\subsection{Extinction in the BLR}\label{sec_extinction}

In virial, SE \mbh\ estimators, either monochromatic or line luminosity (e.g., $\lambda L_\lambda [5100\,{\rm \AA}]$ or \LHa) are used as a probe of the BLR radius. These prescriptions are fundamentally based on RM studies, in which the targets are assumed to be unobscured Type 1 AGNs. However, in the presence of dust extinction along the line of sight, a correction should be applied to the observed line luminosities. In previous work, \citet[][henceforth C20]{Caglar} demonstrated that applying extinction correction reduces $\delm$ by $\sim$0.3-0.4 dex for their sample of 10 Type 1 and 3 Type 2 AGNs. Here, we remind readers that $\delm$ corresponds to the difference in \mbh\ between the SE measurements and the \sig\-based measurements.

In Section \ref{sec:blr_ext}, we discuss the importance of extinction correction for the BLR-based estimates of $\mbh$. We use intrinsic ultra-hard X-ray luminosities to obtain extinction-corrected H$\alpha$ luminosities in the BLR, assuming that the BLR extinction of H$\alpha$ emission is purely due to attenuation by dust \citep{Taro}:

\begin{equation}\label{eq:LHa_corr_Lx}
\log \LHacorr = 1.117\times  \log (L^{\rm int}_{\rm 14-150\,\kev}) - 6.61 \, \: \: \: \: \ergs,
\end{equation}

where $L^{\rm int}_{\rm 14-150\,\kev}$ is the intrinsic X-ray luminosity integrated over the 14--150\,\kev\ energy range and \LHacorr\ is the intrinsic (i.e., extinction-corrected) broad H$\alpha$ luminosity. Here, we note that we adopt the updated $\log L_{14-150\,\kev} - \log \LHa$ correlation parameters (T. Shimizu, private communication) and that this correction introduces an additional systematic uncertainty of $\sim$0.2 dex to the associated \mbh\ estimates (i.e., through Eq.~\ref{eq:mbh_ha}). 

However, since we do not have intrinsic ultra-hard X-ray luminosities in the 14--150\,\kev\ energy band ($\log (L^{\rm int}_{\rm 14-150\,\kev}/\ergs)$) for our bonus sample of 55 AGNs, we will use the observed ultra-hard X-ray luminosities in the 14--195\,\kev\ energy band ($\log (L^{\rm obs}_{\rm 14-195\,\kev}/\ergs)$) as alternatives for $\log (L^{\rm int}_{\rm 14-150\,\kev}/\ergs)$ estimates (see Appendix \ref{sec:LXX}). 

Next, we used the observed and corrected H$\alpha$ luminosities to estimate the level of optical extinction ($A_\mathrm{H \alpha}$) for each source. This is done by deriving the extinction, in magnitudes, affecting H$\alpha$, 

\begin{equation}\label{eq:A_Ha}
A_\mathrm{H \alpha} = 2.5 \times (\log \LHacorr - \log \LHaobs) \, \: \: \: \: \mathrm{mag} ,
\end{equation}

and then deriving the extinction at any wavelength $\lambda$ following the empirically determined extinction law of \cite{Wild}:

\begin{equation}\label{eq:A_lambda}
\frac{A_\mathrm{\lambda}}{A_V} = 0.6\left(\frac{\lambda}{5500\,\mathrm{\AA}}\right)^{-1.3} +0.4\left(\frac{\lambda}{5500\, \mathrm{\AA}}\right)^{-0.7} \, . 
\end{equation}

This extinction law is particularly appropriate for AGNs with a large dust reservoir \citep[e.g.,][]{Wild,Allan}. Here, we note that deriving the extinction in the BLR cannot be done by simple Balmer decrement method (i.e. \ha/\hb\/), as the photo-ionization models predict a wide range of theoretical BLR line ratios for AGNs depending on their BLR conditions \citep{Allan}. Finally, we correct the \mbh\ estimates reported by \citet{Mejia_Broadlines} for the BLR extinction (see Appendix \ref{sec:comparison_extinction_correction_mbh} for a discussion of the difference between extinction-corrected/uncorrected estimates).

\subsection{Eddington Ratio and Accretion Rates}

In order to estimate the Eddington ratios (\eddr), we follow the same approach used in BASS DR1 \citep{2017Kossbass}. First, we convert the intrinsic, absorption-corrected hard X-ray luminosities (\Lhardint) to bolometric luminosities (\Lbol) using a universal bolometric correction, that is $\Lbol = 20 \times \Lhardint$. 
Although this simple bolometric correction may carry significant uncertainties (i.e., $\approx20_{-10}^{+60}$) and likely depends on various AGN properties \cite[see e.g.,][]{Marconi2004,2009Vasudevan}, we note that it was shown to be fairly constant for low luminosity AGNs ($\log (\Lhard/\ergpersec)\lesssim 45$; see, e.g., the study of \citealt{2020Duras} which relies on the Swift-BAT AGN sample). 
Therefore, the fact that the majority of our sample is dominated by such low-luminosity AGNs further justifies the use of a universal bolometric correction. 

The Eddington luminosities ($L_{\rm Edd}$) of our sources are calculated as $L_{\rm Edd} = 1.26 \times 10^{38}\, \mbh/\Msun$, which is appropriate for a pure hydrogen gas. 
Finally, the Eddington ratios are calculated following $\eddr\equiv \Lbol/L_{\rm Edd}$.
We emphasize that the large uncertainty in $L_\mathrm{Bol}$ ($\sim$0.3-0.6 dex) and $\mbh$ ($\sim$0.4 dex) contributes to the (systematic) uncertainty of the Eddington ratio, which is likely ${\gtrapprox}$0.7 dex in total \citep[e.g.,][]{2003Bian,2012Marinucci}. We also estimate the physical accretion rates ($\dot{M}$) that power the AGNs in our sample, through $\dot{M} = \Lbol / \left(\eta c^{2} \right)$, assuming a standard radiative efficiency of $\eta = 0.1$.

\section{Results and Discussion}\label{sec:results}

In this work, we measured \sigs\ of broad-line Type 1 AGNs from the BASS DR2 sample. In the top-left panel of Figure \ref{fig_sig_hist}, we present an overview of our sample of AGNs for successful (173) and failed (68) \sig\ fit results. The majority of \sigs\ measurements are obtained from the \cat spectral region (167 successful and 74 failed fits), but whenever available, we provide the resulting \sigs\ measurements from the \cahk and \mgi\ absorption lines (113 successful and 116 failed fits). 
There are also 12 missing fits (see the top-right panel of Figure \ref{fig_sig_hist}).
For eight of these (BAT IDs: 184, 301, 558, 607, 631, 680, 1046, 1142) we adopt \sigs\ measurements from the LLAMA study by C20.
The remaining five cases lack the appropriate spectral coverage due to the BASS observational setup available at the time of writing. 

In what follows, we discuss the issues related to sample (in)completeness; present a comparison between the \sigb\ and \sigr\ measurements, explore the systematic uncertainties caused by aperture size, and show a direct comparison between our \sig\ measurements and those available from other surveys. 
We then use BASS archival data to address the possible reasons for the offset from the $\mbh - \sig$ relation by looking into trends with key parameters such as BLR extinction, redshift, intrinsic X-ray luminosity, and Eddington ratio. 
Finally, we present the resulting $\mbh - \sig$ relation for our sample of Type 1 AGNs. 

\subsection{Stellar Velocity Dispersion Measurements}

\subsubsection{Sample (in-)completeness}\label{sec:sample_comp}

Thanks to the high-resolution observations with instruments such as VLT/X-shooter, SOAR/Goodman, and Palomar/DBSP, our sample is free of biases caused by insufficient spectral resolution. The spectral setups used for this work provide an instrumental resolution of $\sigma_\mathrm{inst} = 19 - 27\,\kms$, which allows us to measure \sigs\/ of intermediate SMBH hosting AGNs (\mbh\/ $<$ 10$^{6}$ \Msun\/). A limited number of our AGNs have been observed with lower spectral resolution setups, as mentioned in Section \ref{sec:sigs_est}, but we emphasize that the low-resolution were not used in \sigs\ measurements in order to avoid possible biases caused by resolution insufficiency.

We remind the reader that we fit a total number of 240 Type 1 AGN spectra for two distinct spectral regions ($3880-5500$\,\AA\ and $8350-8730$\,\AA, whenever possible). 
These fits yield independent \sigs\ measurements based on the \cahk and \mgi\ features, and on the \cat features, respectively. We refer to these independent \sigs\ measurements as \sigb\ and \sigr\ (again, respectively).

Table \ref{table_mainresult} lists the resulting \sig\ measurements from both spectral regions together with the corresponding quality flags.
We obtained at least one \sig\ measurement for 173 AGNs with small uncertainties. We had 67 failed attempts (see Table \ref{table_mainresult_fail}). For the 173 successful fits, we flag 128 \sigs\ measurements as good, 35 as acceptable, and 10 as unaccepted fits (quality flags 0, 1, and 2, respectively). 
We additionally fit 48 duplicate spectra observed with other instruments, yielding 28 successful and 20 failed fits. 
The top-right panel of Figure \ref{fig_sig_hist} presents the main reasons for failed \sig\ measurements, including: 
strong AGN emission features ($\sim$64\% of the failed fits); insufficient SNR ($\sim$16\%); insufficient spectral resolution ($\sim$14\%); and high Galactic extinction ($\sim$6\%). In addition to these, we also compare our successful and failed \sigs\ measurements with various AGN properties in Appendix \ref{sec:various_AGN_properties}. Please see Appendix \ref{sec:various_AGN_properties} for further discussion. 

In the bottom panel of Figure \ref{fig_sig_hist}, we present the distributions of \sigs\ measurements and their corresponding errors. 
Looking at the two types of \sigs\ measurements, our \sigr\ measurements are in the range  $73 \leq \sigr \leq 278 \, \kms$, with a median of $145\pm7\,\kms$, while the \sigb\ measurements are in the range $82 \leq \sigb \leq 272 \, \kms$ with a median of $143\pm7\,\kms$. We compare \sigr\ and \sigb, for the AGNs for which both types of measurements are available, in the left panel of Figure \ref{fig:all_sigma}.
Clearly, the two types of measurements are in excellent agreement, as is supported by a Spearman correlation test ($\rho = 0.98\pm{0.01}$, $p\ll0.01$)\footnote{Throughout this paper, p-values ($p$) are given in three different thresholds (0.01, 0.05 and 0.1. However, we note that these threshold values are represented as upper and lower limits; therefore, the values can be much larger or smaller than reported threshold values. We also note that the Monte-Carlo-based bootstrapping method is used to estimate the uncertainty in the Spearman rank correlation coefficient \citep{2014Curran}.}. The average offset between the two types of \sig\ measurements is small $\langle \log(\sigr)-\log(\sigb)\rangle = 0.002\pm{0.001}$ dex, and the scatter around the 1:1 relation is 0.027 dex. 
This small level of scatter is probably explained by the somewhat different stellar populations that dominate the absorption features in these spectral regions \citep[see e.g.,][]{Riffel2015}.

\startlongtable
\begin{deluxetable}{rlccccccccc}
\tabletypesize{\scriptsize}
\label{table_mainresult}
\tablecaption{Stellar velocity dispersions and key derived properties for our AGNs.}
\tablehead{
\colhead{BAT ID} & \colhead{Galaxy Name} & \colhead{\sigb} & \colhead{flag$_{\rm B}$} & \colhead{\sigr} & \colhead{flag$_{\rm R}$} & \colhead{$\av$} & \colhead{\eddr} & \colhead{\eddrc} & \colhead{$\dot{M}$} & \colhead{Instrument} \\
\nocolhead{} & \nocolhead{} & \colhead{(\kms)} & \nocolhead{} & \colhead{(\kms)} & \nocolhead{} & \colhead{(mag)} & \nocolhead{} & \nocolhead{} & \colhead{($M_\odot\,{\rm yr}^{-1}$)} & \nocolhead{}\\
\colhead{1} & \colhead{2} & \colhead{3} & \colhead{4} & \colhead{5} & \colhead{6} & \colhead{7} & \colhead{8} & \colhead{9} & \colhead{10} & \colhead{11}
}
\startdata
3 & NGC7811 &   88$\pm$14 & 0 & 91$\pm$8 & 0 & 0.131 & -0.924 & -0.943 &   0.022 &  Palomar/DBSP\\                  &  &  &  &  &  &  &  &  &  &  \\
34 & UGC524 & 156$\pm$5 & 0 &   157$\pm$4 & 0 & 1.586 & -0.853 & -1.146 & 0.043 &  Palomar/DBSP\\                   &  &  &  &  &  &  &  &  &  &  \\
43 & Mrk352 & 97$\pm$6 & 0 &   95$\pm$8 & 0 & 0.207 & -1.582 & -1.621 & 0.021 &  Palomar/DBSP\\                     &  &  &  &  &  &  &  &  &  &  \\
45 & LEDA 1075692 &   195$\pm$7 & 0 & 196$\pm$6 & 0 & 2.668 & -1.081 & -1.564 &   0.139 & VLT/X-Shooter\\ &  &  &  &  &  &  &  &  &  &  \\
51 & RBS149 &  & 9 & 134$\pm$28 & 2 & 0 &   -0.939 & -0.939 & 0.168 &    Palomar/DBSP\\                         &  &  &  &  &  &  &  &  &  &  \\
52 & HE0103-3447 &  & 9 & 182$\pm$21 & 2 & 0 &   -1.404 & -1.404 & 0.155 & VLT/X-Shooter\\                      &  &  &  &  &  &  &  &  &  &  \\
60 & Mrk 975 &  & 9 & 149$\pm$9 & 1 & 0.146   & -1.127 & -1.152 & 0.112 & Keck/LRIS\\                           &  &  &  &  &  &  &  &  &  &  \\
61 & Mrk1152 & 168$\pm$7 & 0   & 170$\pm$5 & 0 & 1.78 & -1.07 & -1.39 & 0.221 &  Palomar/DBSP\\                     &  &  &  &  &  &  &  &  &  &  \\
73 & Fairall9 &  & 9 & 219$\pm$14 & 1 & 0 &   -1.138 & -1.138 & 0.336 & VLT/X-Shooter\\                         &  &  &  &  &  &  &  &  &  &  \\
77 & Mrk359 & 90$\pm$15 & 2 &   101$\pm$10 & 2 & 0 & -0.372 & -0.372 & 0.012 &   VLT/X-Shooter \\
\enddata
\tablecomments{We list the columns in this table as follows. (1) Catalog ID from the 105-month SWIFT-BAT survey. (2) Host galaxy. (3) Stellar velocity dispersion measurement from the \CaHK \& \MgI region (3880--5550~\AA). (4) The quality flag for \sigb\ fit. (5) Stellar velocity dispersion measurement from the Calcium Triplet region (CaT, 8350--8750~\AA). (6) Quality flag for \sigr\ fit. (7) Dust extinction in the BLR. (8) The extinction-uncorrected Eddington ratio. (9) The extinction-corrected Eddington ratio. (10) Accretion rate. (11) The instrument used for the observation. \\
(A portion of the table is shown here for visual guidance and the entire table can be found in machine-readable form.) }
\end{deluxetable}

\startlongtable
\begin{deluxetable}{rlcccccccc}
\tabletypesize{\scriptsize}
\label{table_mainresult_fail}
\tablecaption{Stellar velocity dispersions failures}
\tablehead{
\colhead{BAT ID} & \colhead{Galaxy Name} & \colhead{Reason} & \colhead{$z$} & \colhead{E(B-V)} & \colhead{$L_{\rm Bol}$} & \colhead{Seyfert Type} & \colhead{Instrument} \\
\nocolhead{} & \nocolhead{} & \nocolhead{} &\nocolhead{} & \colhead{(mag)} &\colhead{(\ergpersec)} & \nocolhead{} & \nocolhead{}  \\
\colhead{1} & \colhead{2} & \colhead{3} & \colhead{4} & \colhead{5} & \colhead{6} & \colhead{7} & \colhead{8}}
\startdata
6 & Mrk335 & AGN & 0.0259 & 0.0354 & 44.29 & Sy1.2 &  Palomar/DBSP \\
22 & Z535-12 & AGN & 0.0476 & 0.0666 & 44.84 & Sy1.2 &  Palomar/DBSP \\
36 & Mrk1148 & AGN & 0.064 & 0.057 & 45.35 & Sy1.5 &  Palomar/DBSP \\
78 & MCG-3-4-72 & AGN & 0.0429 & 0.0188 & 44.84 & Sy1 &  Palomar/DBSP \\
113 & LEDA138501 & AGN & 0.0497 & 0.1628 & 45.42 & Sy1 &  Palomar/DBSP \\
122 & 2MASXJ02223523+2508143 & Resolution & 0.0616 & 0.08 & 45.05 & Sy1 &  Palomar/DBSP \\
130 & Mrk1044 & AGN & 0.0173 & 0.0334 & 43.81 & Sy1 & VLT/X-Shooter \\
143 & Rhs15 & lowSN & 0.0697 & 0.0662 & 44.98 & Sy1 &  Palomar/DBSP \\
147 & Q0241+622 & AGN & 0.0447 & 0.7427 & 45.54 & Sy1.2 &  Palomar/DBSP \\
161 & 2MASXJ02593756+4417180 & Resolution & 0.0313 & 0.2206 & 44.39 & Sy1 &  Palomar/DBSP \\
\enddata
\tablecomments{Column descriptions: (1)  Catalog ID from the SWIFT-BAT survey, (2) Host galaxy, (3) Reason for failure. AGN: spectra were dominated by AGN emission lines contaminating the absorption lines, lowSN: low signal-to-noise ratio, Resolution: no suitable high-resolution spectra were available, and GalExt: high Galactic extinction. (4) Redshift. (5) Interstellar reddening. (6) The bolometric AGN luminosity. (7) AGN type based on optical spectroscopy (8) Best available spectra from DR2. See \citet{Koss_DR2_catalog,Mejia_Broadlines} for more details on redshift and Seyfert types.\\
(A portion of the table is shown here for visual guidance and the entire table can be found in machine-readable form.) }
\end{deluxetable}

\subsubsection{Comparison With Other Measurements}

\begin{figure}[!ht]
    \centering
    \includegraphics[width=0.494\textwidth]{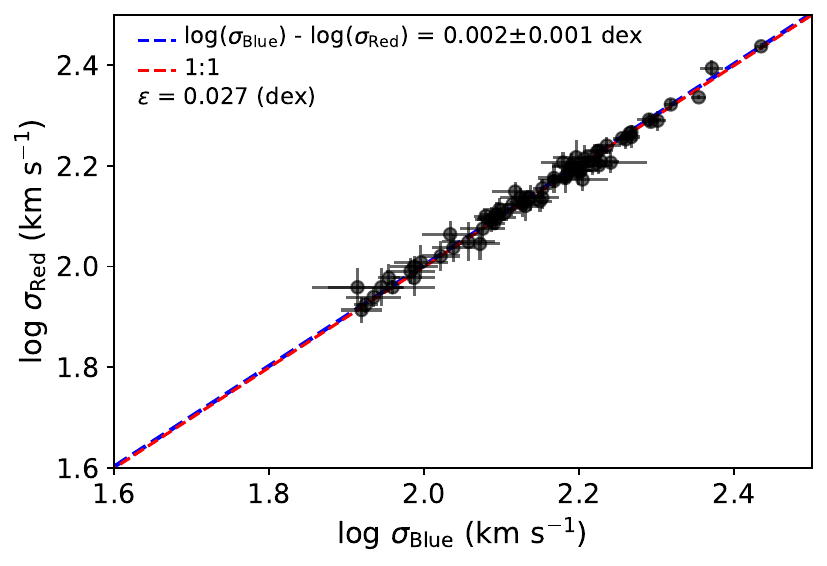}
    \includegraphics[width=0.494\textwidth]{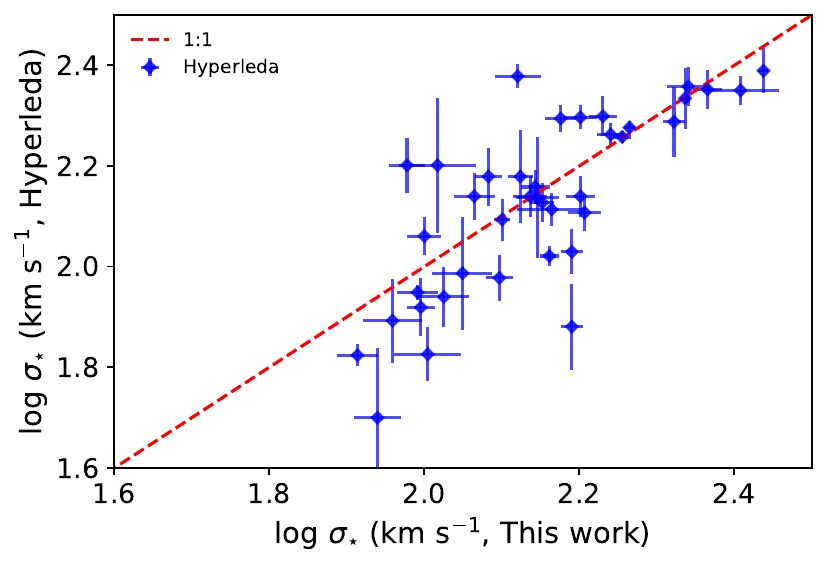}
    \caption{\textbf{Left:} Comparison between \sigb\ and \sigr\ measurements. The blue dashed line represents the difference in both measurements. The red dashed lines represent 1:1 lines in both panels. \textbf{Right:} comparison between our best fitting \sig\ results versus the \sig\ results in the literature. }
    \label{fig:all_sigma}
\end{figure}

We compare our best \sigs\ measurements with  literature measurements from the HyperLeda \sigs\ catalog \citep{Leda}, which contains a total of nearly of 40,000 \sigs\ measurements for more than 29000 objects. We find a total of 39 \sigs\ measurements, drawn from nine studies \citep{NelsonW,Terlevich,Oliva,Oliva2,Wegner2003,GR2005,Greene2006,Ho2009,Cappellari2013}. In the right panel of Figure \ref{fig:all_sigma}, we present a comparison between our best \sig\ measurements and the corresponding measurements from HyperLeda. The difference in median between our \sig\ measurements and those of HyperLeda is $\sim$0.006 dex, which could be caused by aperture differences. Interestingly, our \sig\ measurement uncertainties (a median value of 7 \kms\/) are typically lower than those reported in HyperLeda (a median value of 13 \kms\/). Here, we note that we define the best \sigs\ measurements using two conditions; I) the ones with better fit quality among \sigb\ and \sigr\ fits and II) the ones with smallest uncertainty in \sigs .

\subsection{The $M_\mathrm{BH} - \sigma_\mathrm{\star}$ Relation of BAT Type 1 AGNs}

Our sample and measurements enable one of the largest investigations of the $\mbh - \sig$ relation for Type 1 AGNs. We fit our $\log\sigs$ and $\log\mbh$ measurements with a linear relation using the bivariate correlated errors and the intrinsic scatter method, which takes into account the measurement errors in both variables (i.e., $X$ and $Y$ axes; \citealt{1996Akritas,2012Nemmen}). The linear regression was performed using the Y/X method, where the slope and intercept can vary. To fit the $\mbh - \sig$ relation, we use a single power law function as expressed in the following equation:

\begin{equation}
\log{(M_{BH}/M_{\odot})} = \alpha + \beta \log{\left( \frac{\sigma_{\star}}{\sigma_{0}} \right)},
\end{equation}

where $\alpha$ is the intercept, $\beta$ is the slope and $\sigma_{0}$ is the normalization coefficient of 200 \kms\/. We then performed the linear regression fits for 4 different data sets, as follows: 
(1) all our AGNs, without extinction corrections (154 sources; DS1 hereafter); 
(2) AGNs with negligible extinction, $\av = 0$ mag (45 sources; DS2 hereafter); 
(3) AGNs with limited extinction, $\av < $ 1 mag (99 sources; DS3 hereafter); 
and (4) all AGNs with extinction corrections (154 sources; DS4 hereafter). 
In Figure \ref{fig_m_sigma_for_DS}, we present the resulting $\mbh - \sigs$ relations for each of our data sets, while Table \ref{table_m_sigma_fit_results} lists the best-fit intercepts ($\alpha$), slopes ($\beta$), and intrinsic scatters ($\epsilon$) derived for each of these data sets.

\begin{figure*}[bt!]
    \centering
    \includegraphics[width=0.70\textwidth]{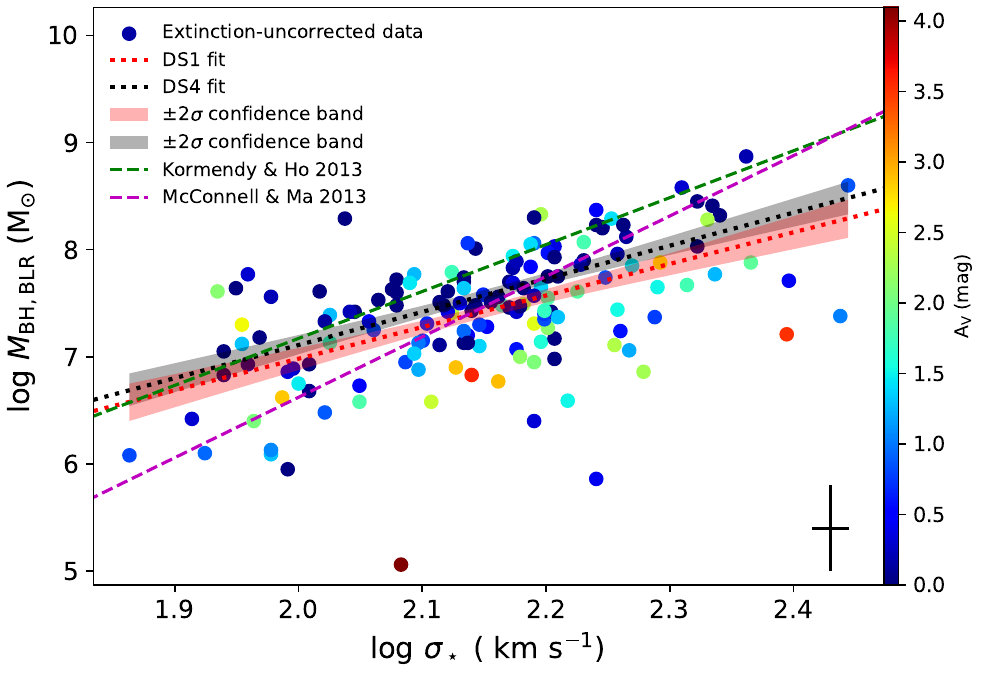}
    \caption{The $\mbh - \sig$ relation of 154 Type 1 AGNs for both extinction-uncorrected and -corrected data sets. The red and black dotted lines represent the resulting $\mbh - \sig$ fits for the DS1 and DS4 subsamples, respectively. We show the median uncertainty in \sigs\ and \mbh\ as a black plus sign for visual aid.}
    \label{fig:starofthepaper}
\end{figure*}

Looking into our best-fit fitting parameters, we note a few key results.
First, the slope of DS1 ($2.95\pm0.41$) is shallower than the slopes found for the other data sets (DS2: $3.21\pm0.66$, DS3: $3.44\pm0.48$, and DS4: $3.09\pm0.39$). This result again implies that the BLR extinction might be somewhat responsible for flattening the slope. However, we cannot statistically confirm this, since uncertainties in the slopes are quite large for our data sets. Second, the range of slopes derived for our BASS sample, $2.54 \leq \beta \leq 3.92$ is consistent with what is found in previous studies. Specifically, our results are consistent with those presented by C20, which reports a slope of $\beta = 3.38\pm{0.65}$, an intercept of $\alpha = 8.14\pm0.20$ and an intrinsic scatter of $\epsilon = 0.32\pm0.06$ for the LLAMA sample. 
Our slopes are also consistent with the slope reported by \cite{Woo13} for a sample with RM measurements ($3.46\pm 0.61$). 
On the other hand, the slope of $4.38\pm 0.29$ reported by KH13 is not consistent with the slopes of our data sets. 
Moreover, none of the slopes we derive is consistent with the steep slope of $5.64\pm 0.32$ reported by MM13 (which included bright central cluster galaxies). The more recent study by \citet{Bennert} reported a slope of $3.89\pm0.53$ for 29 RM AGNs and a slope of $4.55\pm0.29$ for 51 inactive galaxies. Compared with these, our results for the BASS AGNs are consistent with the RM AGN sample of \citet{Bennert} slope (within uncertainties) but are inconsistent with the slope found for inactive galaxies. Thus, our analysis strengthens the evidence that AGNs show a shallower $\mbh - \sigs$ relation compared to inactive galaxies. 

In Figure \ref{fig:starofthepaper}, we present the best-fitting $\mbh - \sigs$ relations for the DS1 and DS4 data sets (along with the corresponding $\pm 2\sigma$ confidence ranges (also see Appendix \ref{sec:M_sigma_plots_DS1234}). Additionally, we compare our results with other $\mbh - \sigs$ relations reported by KH13 and MM13. A significant fraction of AGNs is found to be below the canonical $\mbh - \sigs$ relation reported by KH13. This discrepancy appears to increase as the extinction in broad line regions (BLRs) increases, which is discussed further in Section \ref{sec:blr_ext}. The presence of shallower slopes and large scatter in the low-mass end of the \mbh\ - \sigs\ relation can be seen from the Figure, therefore, this further pushes us to discuss the fundamental differences between AGNs and inactive galaxy samples causing this discrepancy.  

\begin{table}[hbt!]
\begin{center}
\caption{The $\mbh - \sig $ relation results for our data sets}\label{table_m_sigma_fit_results}
\begin{tabular}{lcccc}
\hline\hline
Sub-sample\footnote{As suggested by C20, NGC 7213 is removed from the data sets due to its unreliable \mbh\ measurement since this galaxy hosts a low-ionization nuclear emission-line region.} & Number &  $\alpha$  & $\beta$ & $\epsilon$ \\
\hline
All (no corr) & 154 & 7.87$\pm{0.07}$ & 2.95 $\pm{0.41}$ & 0.24$\pm{0.05}$ \\
$\av$ $=$ 0 & 45 & 8.12$\pm{0.10}$ & 3.21 $\pm{0.66}$ & 0.38$\pm{0.06}$ \\
$\av$ $<$ 1 & 99 & 8.05$\pm{0.09}$ & 3.44 $\pm{0.48}$ & 0.25$\pm{0.04}$ \\
$\av$ (corrected) & 154  & 8.04$\pm{0.07}$ & 3.09 $\pm{0.39}$ & 0.33$\pm{0.06}$ \\
\hline\hline

\end{tabular}
\end{center}
\end{table}

\subsection{Understanding the differences between AGNs and inactive galaxies}

We have demonstrated that our sample of Type 1 AGNs shows significantly shallower $\mbh-\sigs$ slopes relative to the canonical relation determined for inactive galaxies. This may be driven by multiple effects, related both to (host) galaxy evolution and BLR structure, as discussed below.

First, one can postulate that AGNs must follow the same $\mbh-\sigs$ relation as inactive galaxies, in which case the observed discrepancy may be attributed to variations in the BLR geometry \citep{Onken04}. 
We recall that efforts to obtain an average $f$ factor have been limited to only a few dozen AGNs with a relatively narrow range of \mbh\ and \eddr. 

Moreover, some studies suggest that the BLR geometry (i.e., as encoded in the $f$-factor) may depend on some fundamental BH properties, including both observed trends between $f$ and $FWHM$, \mbh\ and/or \eddr\ \cite[e.g.,][and references therein]{2017StorchiBergmann,2018MJNatAs}, and disk-wind models that anticipate such trends \cite[e.g.,][]{2004Proga}.

Second, since \mbh\ is estimated by different methods for AGNs and inactive galaxies, different selection effects and biases may affect these two kinds of samples. As discussed by \citet{Bernardi2007,Shankar2016}, a resolution-dependent bias affects dynamical \mbh\ determinations in inactive galaxies, which does not affect AGNs\footnote{See also \cite{vdBosch2015} for further discussion of possible biases regarding inactive galaxies.}. On the other hand, the RM AGN samples based on which the best \mbh\ measurements are obtained, and the SE method is based, may also be biased.
In particular, most RM efforts have been focused on low-redshift, low-to-medium luminosity AGNs, where sufficient variability can be expected and where emission line time lags can be more robustly monitored (but see exceptions in, e.g., \citealt{Lira2018}). 
In addition to the challenge of measuring \sigs, and thus inferring the \mbh\ scaling ($f$) in such luminous sources \cite[e.g.,][]{Grier2013}, it is also possible that such AGNs may not be representative of highly luminous AGNs like those probed by BASS or by high-redshift surveys (see detailed discussion of luminosity-related biases in, e.g., \citealt{2008Shen}).
These difficulties add to other issues concerning which broad emission line, and which line profile measurement, best probe the virialized BLR motion \citep[e.g., see ][]{2004Peterson,2006Collin,2020DallaBonta}, how to inter-calibrate SE prescriptions based on various emission lines \cite[e.g.,][]{Shen2012,2017Park,2018MJNatAs,2018MJ_Blackholes,2020DallaBonta}, as well as BLR extinction (as shown in this paper). 

Third, significant differences in the kind of SMBHs probed through our AGN sample and the literature inactive galaxy sample, as reflected in their different \mbh\ distributions, might also play a role in the slope discrepancy. The extinction-uncorrected \mbh\ estimates for our AGN sample cover the range $5 \lesssim \log (\mbh/\Msun) \lesssim 8.9$, with a median value of ${\approx}7.5$. 
However, an important fraction of the inactive galaxy sample (35\% of the total KH13 sample) consists of SMBHs with $\log (\mbh/\Msun) > 9$. On the other hand, 31 of our AGNs (20\%) are found to have $\log (\mbh/\Msun)\lesssim 7$, compared with only two such SMBHs (4\%) among the KH13 sample of inactive galaxies \citep[see][ for more detail]{Bennert}. 
Thus, the high-mass regime is significantly over-represented by the inactive galaxies sample with \sigs\ measurements (or, alternatively, under-represented in our AGN sample; see \citealt{Ananna_DR2_XLF_BHMF_ERDF} for a detailed census of \mbh\ distributions among BAT AGNs). 
The lack of \sigs\ measurements in the high-\mbh\ regime might flatten the slope of the relation for AGN samples, whereas it might result in a steeper relation for inactive galaxy samples if those lack low-\mbh\ systems. The latter may reflect, again, known biases in our current ability to measure \sigs\ in inactive galaxies in the local Universe (see above).

The differences in \mbh\ distributions between the active and inactive galaxy samples could also reflect deeper differences in the evolutionary paths experienced by the two galaxy populations.  In this context, we note that the majority of elliptical, inactive galaxies (or those with classical bulges) are thought to be the result of a previous major merger \citep[see, e.g.,][]{2004Kormendy,koho13}, while disk-dominated galaxies, or galaxies with pseudo-bulges and/or bars, are thought to be dominated by secular evolution. 
Such galaxies typically host lower-mass SMBHs, and were shown to present significantly larger scatter, and shallower slopes, for their $\mbh - \sigs$ relations \citep[e.g.,][]{Graham2008,2008Hu,Greene10,Ho14}. 
Combined with our findings, it is thus possible that our AGN hosting galaxies tend to be disk-dominated, or to have pseudo-bulges (and/or bars), and to mark evolutionary paths that are different than those of large elliptical (or bulge-dominated) galaxies (see \citealt{2011Koss} for additional evidence for disk dominance among BAT-selected AGNs).

Finally, we note two more subtle issues when considering the $\mbh-\sigs$ relations of active and inactive galaxies.
From a theoretical perspective, the growth of SMBHs and of the bulges that host them does not have to be perfectly synchronized \citep[e.g.,][]{2005HoHo,2006Kim,2012Volonteri,2019Ricarte}, which will introduce additional scatter in the $\mbh - \sigs$ relation. 
Specifically for our findings, as we are focusing on rather powerful AGNs, the SMBHs are experiencing a significant growth episode, which might not be echoed by a corresponding \sigs\ increase, thus leading to an expectation for the systems to grow ``towards'' the canonical $\mbh-\sigs$ relation (in the near cosmic future; see also Section~ \ref{sec_eddr_accr_rate}). From a practical perspective, host galaxy disk contamination can increase the observed \sigs\ by up to $\approx$25\% (due to orientation and/or rotation; see, e.g., \citealt{2013Kang,2014Bellovary,2017Eun};C20). In addition, as suggested by \citep[][]{2013Debattista}, the compression of bulge caused by the disk formation might introduce an increase of 10\% in \sigs\/. Therefore, the increased \sigs\ might cause additional offsets in the $\mbh - \sigs$ relation for galaxies hosting disks.

\subsection{The Offset from the $M_\mathrm{BH}$ - $\sigma_\mathrm{\star}$ relation}

\begin{figure*}
    \centering
    \includegraphics[width=0.489\textwidth]{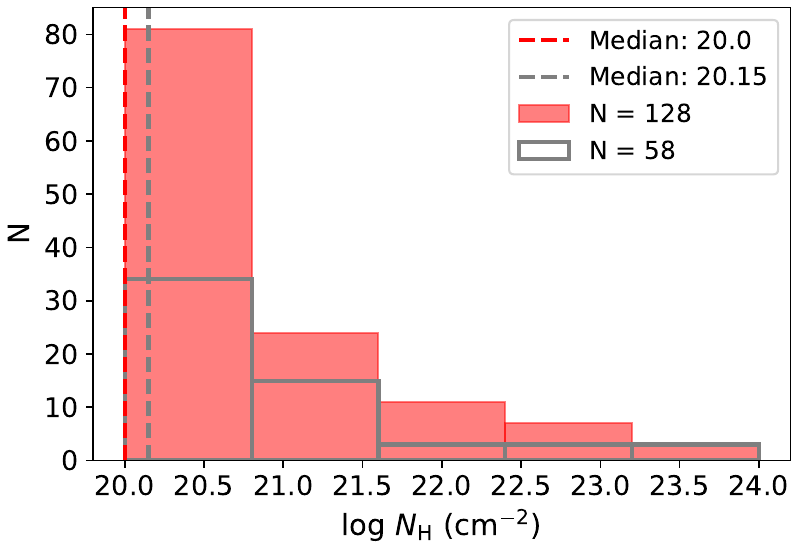}
    \includegraphics[width=0.489\textwidth]{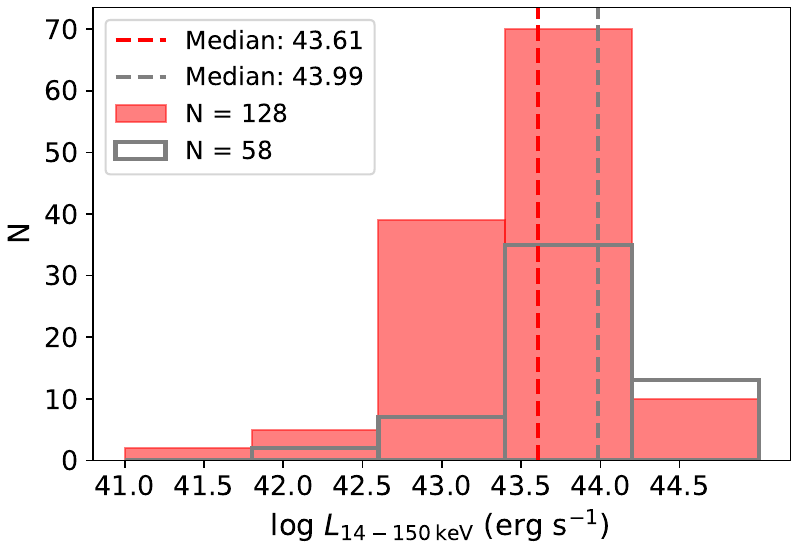}
    \includegraphics[width=0.487\textwidth]{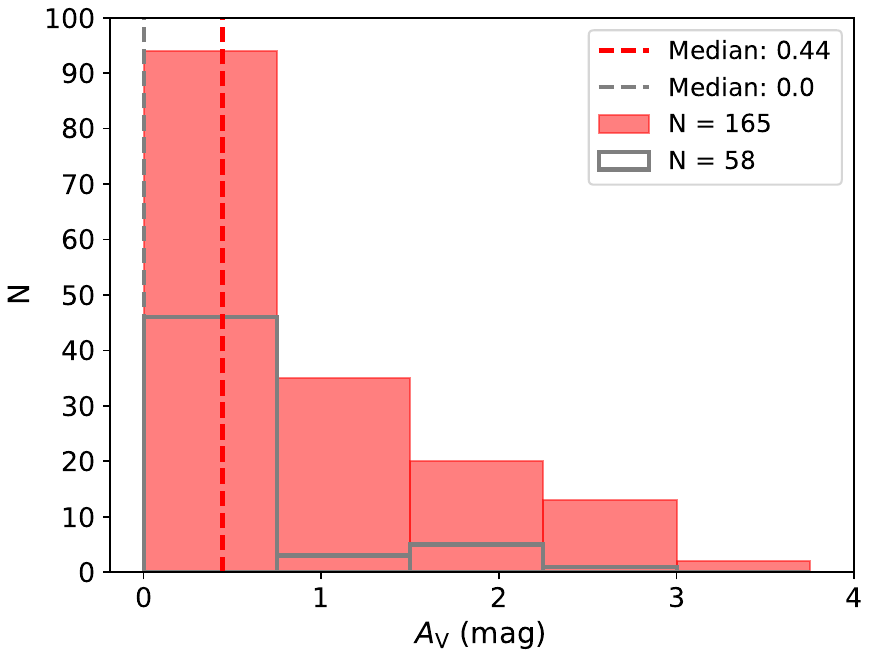}
    \includegraphics[width=0.499\textwidth]{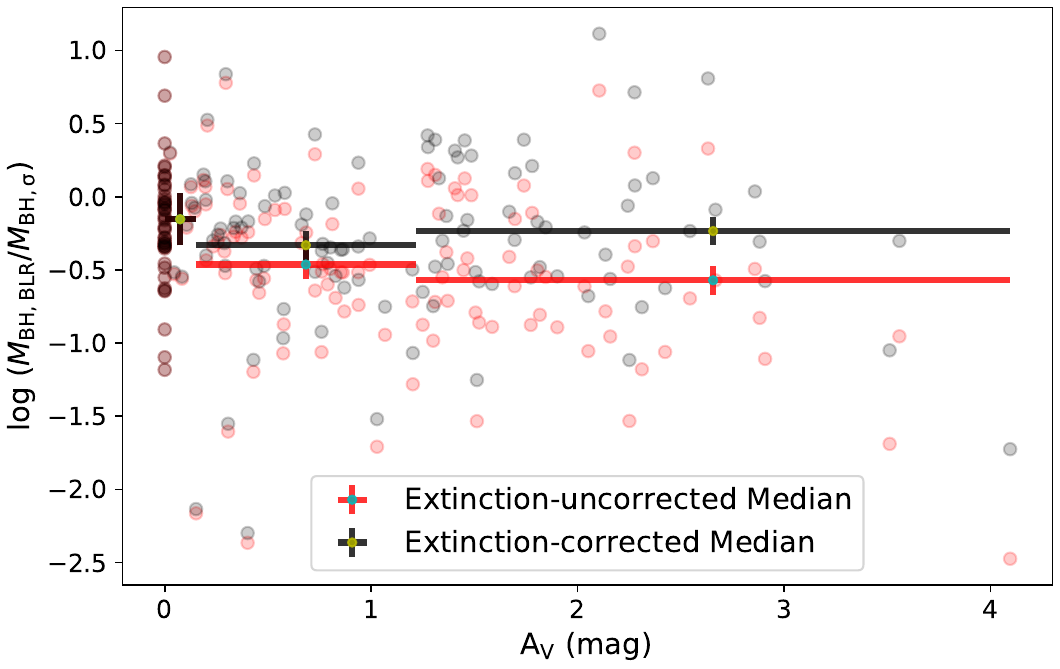}
    \caption{ The distribution of hydrogen column density \textbf{(top left)}, hard X-ray luminosity \textbf{(top right)} and the extinction in the BLR \textbf{(bottom left)}. The red columns represent the presented parameter's distribution for AGNs with successful \sig\ fits, whereas the grey ones represent the ones with failed \sig\ fits. Median values are presented as red and grey dashed vertical lines, respectively. \textbf{Bottom right:} the offset from the $\mbh - \sig$ relation versus the extinction in the BLR for our sample of AGNs.The medians (together with the bin edges and the standard error on the median)} are presented for three bins with equal numbers of data points.
    \label{fig:L150}
\end{figure*}

\subsubsection{The Extinction in the BLR}\label{sec:blr_ext}

We next look into the BLR extinction, as can be determined from the suppression of broad H$\alpha$ line emission for any given ultra-hard X-ray luminosity. 
Assuming such suppression is caused entirely by dust extinction, and that the gas-to-dust ratio is similar to that of the Milky Way, i.e. yielding $\nh/\av = 1.79-2.69 \times 10^{21} \, \cmii$ \citep{Predehl,Nowak}, we expect a maximum $\nh$ threshold of ${\approx}10^{22.3}\,\cmii$ for Type 1 AGNs (i.e., excluding Sy 1.9s with broad \Halpha\ emission-lines) for discriminating X-ray absorbed and unabsorbed sources as described by \citet{Leo1}. 
We note that there are only 10 absorbed AGNs in our sample (see the top left panel of Fig.~ \ref{fig:L150}). The median $\log(\nh/\cmii)$ values for our AGNs with successful and failed \sigs\ fits are 20.0 and 20.15, respectively. The majority of our $\nh$ estimates cluster around $\log(\nh/\cmii) = 20.0$. This value is the upper limit due to Galactic extinction placed by \cite{Ricci17a} for completely unobscured sources, since lower intrinsic $\nh$ values cannot easily be determined. We also stress that we have no Compton-thick sources ($\log[\nh/\cmii] >  24$) in our sample. 

We can thus use the $\log (L^{\rm int}_{\rm 14-150\,\kev}/\ergs)$ as a probe of the extinction-corrected broad $\halpha$ luminosities (\LHacorr). In the top-right panel of Figure \ref{fig:L150}, we report the distribution of $\log (L^{\rm int}_{\rm 14-150\,\kev}/\ergs)$ for AGNs with successful and failed \sigs\ fits resulting in the median values are 43.61 and 43.99, respectively. The difference between the observed and intrinsic $\halpha$ luminosities, accordingly, gives us the X-ray-derived $\halpha$ extinction ($\AHa$) for the BLR (see Equations \ref{eq:LHa_corr_Lx} and \ref{eq:A_Ha}). We note that applying such conversion introduces an average uncertainty of 0.4 dex in \LHa\ estimations \citep{Taro}. 
The bottom-left panel of Figure~ \ref{fig:L150} shows the distribution of $\av$ derived through our approach, which has median and mean $\av$ (see Eq. \ref{eq:A_lambda} for the conversion between $\av$ and $\AHa$) values of 0.44 and 0.84 mag, respectively. 
Importantly, a significant fraction (66\%) of our AGNs has $\av < 1$ mag. We also stress that there are only 22 AGN with $\av > 2$ mag. We note that extinction correction has a very large uncertainty for highly extincted sources, and thus should be used with great caution. Finally, we suggest using near-infrared broad-emission lines for such extreme cases \citep[][]{Ricci_DR2_NIR_Mbh}. 

In the bottom-right panel of Figure~\ref{fig:L150}, we present the offset of our AGNs from the canonical $\mbh - \sig$ relation of KH13, which was defined as $\delm \equiv \log(M_\mathrm{BH,BLR}/M_\mathrm{BH,\sigma})$ in the introduction, versus the BLR extinction estimates. 
We plot these for both the extinction-uncorrected and extinction-corrected data sets. 
Despite the significant scatter in this parameter space diagram, we find a statistically significant anti-correlation, as supported by a formal Spearman correlation test ($\rho = -0.38\pm{0.07}$, $p\ll0.01$) showing that the extinction in the BLR plays a role in the offset. 
Applying the extinction correction reduces \delm\ to some extent, but \delm\ persists across all extinction regimes (see binned data points in Figure \ref{fig:L150}). 
This result is a confirmation of the findings by C20, which used a significantly smaller sample of Swift-BAT detected AGNs with a redshift cutoff of $z< 0.01$ and an ultra-hard X-ray luminosity cutoff of $\log (L_{14-195\,\kev} / \ergs) \geq 42.5 $. 

We emphasize that applying the (uncertain) extinction correction may introduce significant additional uncertainty to \mbh\ estimates (through its dependence on \LHa), for two reasons. First, the extinction corrections themselves are somewhat uncertain. Second, the potential flux variability between the BAT X-ray measurements and the optical spectroscopy can cause an over-correction by up to $\sim$1 dex for some sources. To demonstrate the scope and challenges of the extinction corrections, we consider the individual case of NGC 1365 (BAT 184). The observed broad H$\alpha$ luminosity for this source is $\log (\LHaobs/\ergs) = 39.37$ and the extinction we derive is $\AHa = 3.40$ mag, yielding an extinction-corrected line luminosity of $\log (\LHacorr/\ergs) = 40.73$, i.e. an upward correction of 1.36 dex. Correspondingly, the extinction-corrected $\mbh$ differs from the raw $\mbh$ estimate (i.e., Eq.~\ref{eq:mbh_ha}) by 0.75 dex. Here, we note that NGC 1365 is a well-known changing look AGN \citep[e.g.,][]{2000Risaliti,2009Risaliti,2014Walton,2022Mondal,2022Temple,Ricci23}, therefore, the variability can be somewhat responsible for this discrepancy for such sources.

As shown throughout this paper, we claim that the extinction in the BLR can cause considerable under-estimation of $\mbh$ for highly obscured AGNs unless it is taken into account. This result is also shown by C20 and recent BASS studies \cite[][]{Mejia_Broadlines,Ricci_DR2_NIR_Mbh}. However, applying extinction correction increases the noise in data. Therefore, as proposed by \citet{Ricci_DR2_NIR_Mbh}, we encourage measuring black hole masses using near-infrared broad-emission lines, which are expected to be less affected by dust extinction.

\begin{figure*}
    \centering
    \includegraphics[width=0.494\textwidth]{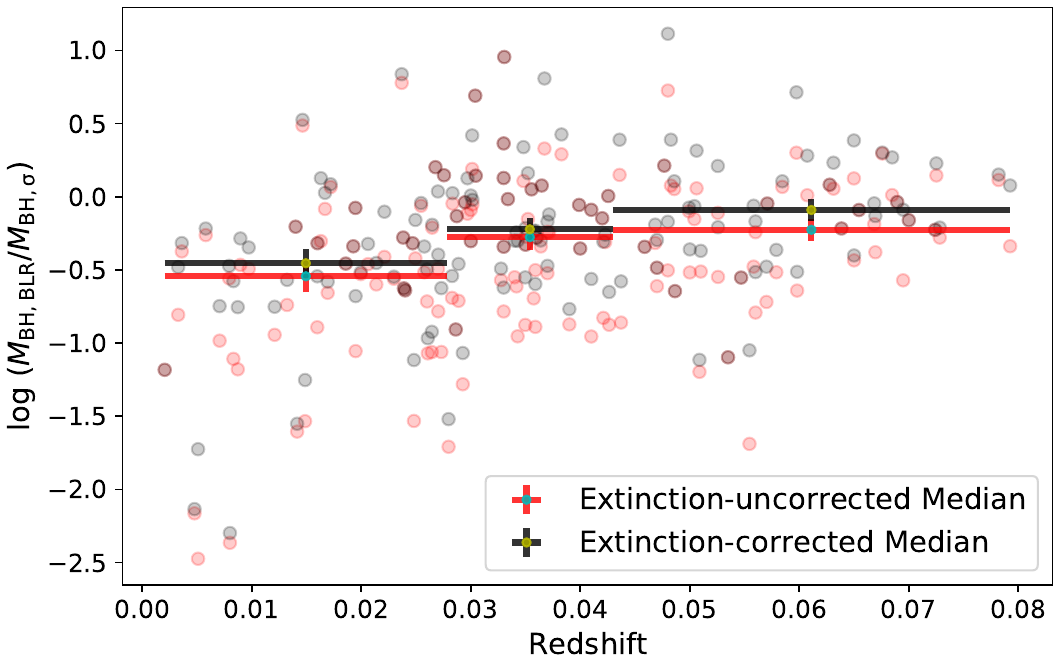}
    \includegraphics[width=0.494\textwidth]{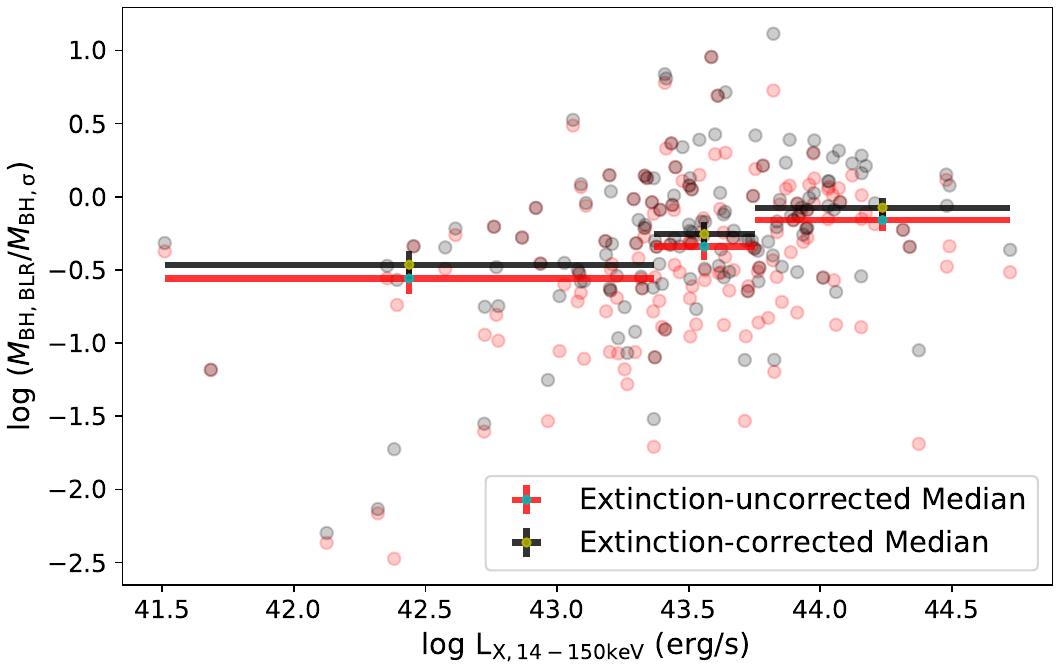}
    \caption{\textbf{Left:} The comparison between \delm\ and redshift. \textbf{Right:} The comparison between the offset from the $\mbh - \sig$ relation and hard X-ray luminosity. The medians (together with the bin edges and the standard error on the median)} are presented for three bins with equal numbers of data points.
    \label{fig:deltaz}
\end{figure*}  

\subsubsection{Redshift and Intrinsic X-ray Luminosity}\label{sec_z_aperture}

In Figure \ref{fig:deltaz}, we present a direct comparison between \delm\ and both $z$ (left panel) and $\log (L^{\rm int}_{\rm 14-150\,\kev}/\ergs)$ (right panel), for both extinction-corrected and uncorrected measurements. 
We see trends of increasing \delm\ with both increasing redshift and luminosity, regardless of the extinction correction. 
For the trend with redshift, a Spearman test results in correlation coefficients of $\rho= 0.33\pm{0.03}$ and $0.36\pm{0.04}$ for the extinction-uncorrected and extinction-corrected data, respectively (with $p\ll0.01$ for both cases). 
For the trend with X-ray luminosity, the corresponding correlations are $\rho=0.32\pm{0.03}$ and $0.38\pm{0.04}$ (with $p\ll0.01$ for both cases). 
Although each of these trends is statistically significant, they are very likely interleaved given the flux-limited nature of the BAT survey. There are two-sided biases here; 1) as redshift increases, the chance of detecting lower luminosity sources by X-ray instruments decreases, 2) the number of X-ray bright AGNs are limited in the nearby universe \cite[e.g.,][]{DaviesLLAMA,2017Caglar}. In fact, $L_\mathrm{H\beta}$, $L_\mathrm{5100}$ and $L_\mathrm{H\alpha}$ are strongly correlated with $\log(1+z)$ ($\rho =$ 0.51$\pm{0.03}$, 0.54$\pm{0.03}$ and 0.66$\pm{0.02}$, respectively) which indicates a strong redshift-luminosity selection bias. We also point out that some contribution to this trend might come from projection effects caused by the limitation of instrumental aperture sizes since the $\mbh - \sig$ relation is assumed to hold at effective radii (see the discussion in Appendix \ref{sec_morp_aperture}). The $\mbh$ estimates from the BLR are not affected by such limitations, since BLR gas resides at sub-parsec scales.

\begin{figure*}[t!]
    \centering
    \includegraphics[width=0.49\textwidth]{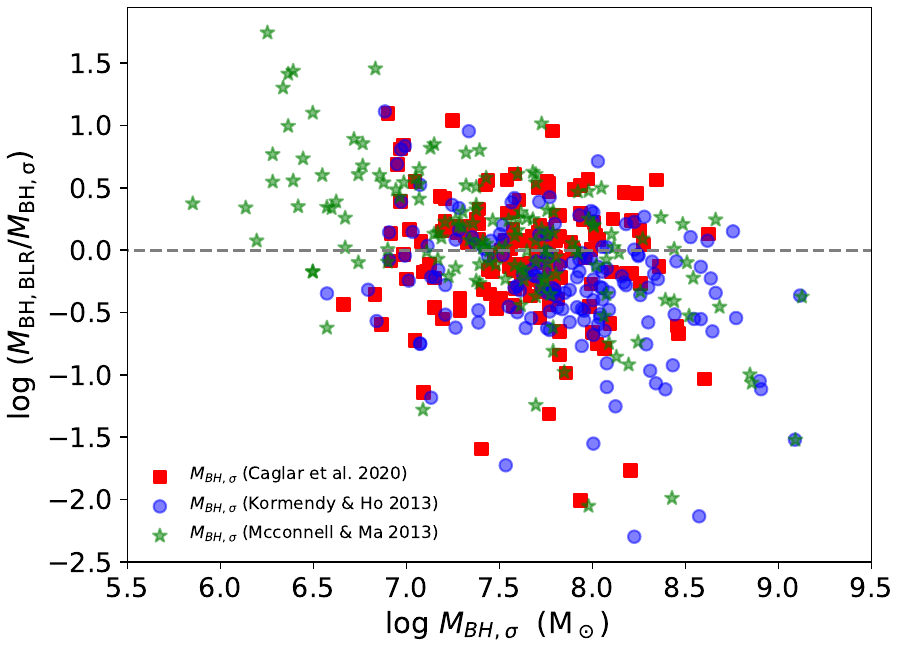}
    \includegraphics[width=0.49\textwidth]{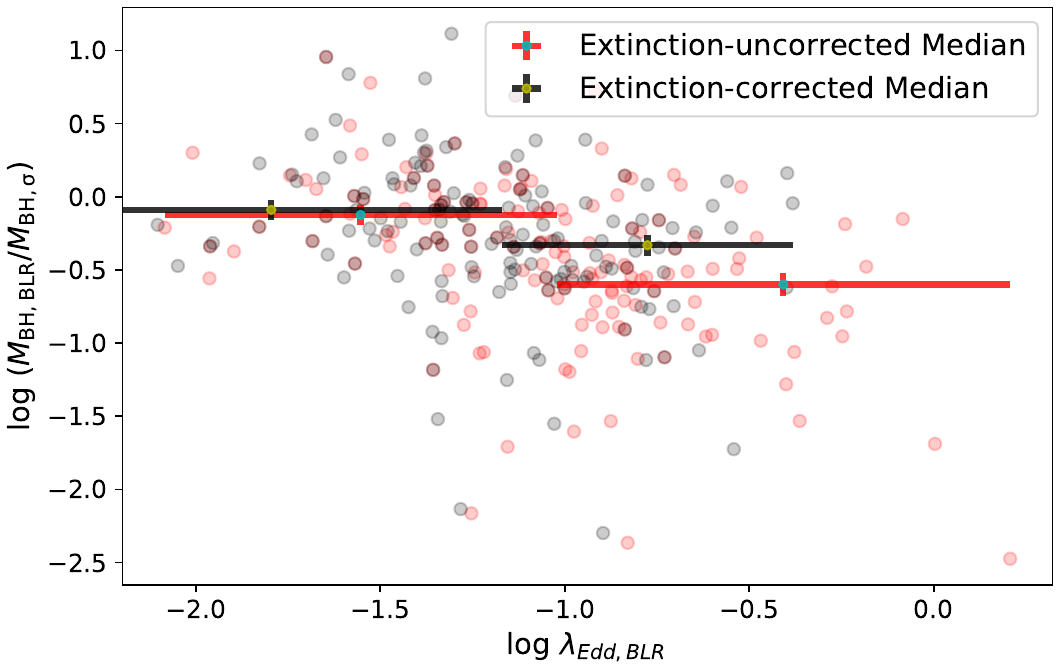}
    \caption{The comparison between the adopted the $\mbh - \sig$ relation as an estimator of M$_\mathrm{BH}$ (left). The \delm\ versus Eddington ratio (right). The medians (together with the bin edges and the standard error on the median)} are presented for two bins with equal numbers of data points.
    \label{fig:kohovcaglar}
\end{figure*}

\subsubsection{Impact of the $M_\mathrm{BH} - \sigma_*$ Relation Used}

The slope of the $\mbh - \sig$ relation may be indicative of the physics behind the AGN-driven feedback mechanism. Specifically, $\mbh \propto \sigma^{4}$ corresponds to momentum-driven feedback while $\mbh \propto \sigma^{5}$ corresponds to energy-driven feedback \cite[e.g.,][]{Silk,King}. 
Previous efforts to determine an (universal) $\mbh - \sig$ relation yielded a wide range of slopes, i.e. $\beta \simeq 3.7-5.6$ for inactive galaxies \citep[e.g.,][]{Ferrarese,Geb,Tremaine,Gultekin,2012Beifiori,Batiste17} and $\beta \simeq 3.4-4.0$ for AGNs \cite[e.g.,][]{Woo13,Woo15}. 
Although most of these studies report a tight relation, with an intrinsic scatter of $\gtrapprox$0.3 dex, the uncertainties on the slope typically exceed $\Delta\beta \simeq 0.3$. AGN samples tend to show both larger uncertainties (due to their smaller size) and flatter slopes \cite[see also][]{SDSSRM}.
Adopting different $\mbh-\sig$ relations naturally results in additional differences in \mbh, ranging from $\sim$0.35 dex for a fiducial $\sig = 150\,\kms$ to over 0.8 dex for the lower and higher ends of the \sigs\ (or \mbh) distribution. 

In the left panel of Figure \ref{fig:kohovcaglar}, we plot $\mbh$ estimates for our BASS sample obtained from three different scaling relations (C20, KH13, and MM13) versus the corresponding \delm (the difference between the extinction-corrected \mbh estimates versus the ones from the adopted $\mbh-\sig$ relation). We note that these three calibrations were derived using different samples of different types of galaxies: the KH13 relation is based on elliptical and classical bulge galaxies, whereas the MM13 sample consists of early and late-type galaxies, as well as BCGs. On the other hand, the C20 sample consists of luminous, hard X-ray-selected local Type 1 AGNs. 
It appears that black hole masses obtained using the $\mbh - \sigma$ relation reported by C20 are relatively closer to the zero point (with a median offset of 0.03$\pm{0.04}$ dex) compared to the ones reported by KH13 (with a median offset of -0.24$\pm{0.05}$ dex) and MM13 (with a median offset of 0.11$\pm{0.05}$ dex). Using our sample of AGNs, we report that the scaling relation by KH13 shows a tendency to overestimate black hole masses, whereas the scaling relation by MM13 tends to underestimate black hole masses. On the other hand, the C20 version of the $\mbh - \sigma$ relation provides a better description for the BASS data set. This result indicates that AGNs might be following a different $\mbh - \sig$ relation.

\subsubsection{The Eddington Ratio and Accretion Rates}\label{sec_eddr_accr_rate}

In this section, we investigate whether \delm\ is correlated with the Eddington ratio. We first note that the extinction-uncorrected Eddington ratio estimates for our sample are in the range $-2.10 < \eddr < 0.21$, with a median value of $\eddr \simeq -1$. 
Only two sources have $\eddr > 0$ (NGC 1365 and PKS0521-36); however, both have high BLR extinction ($\av > 3.5$ mag), and thus their $\mbh$ are under-estimated and, accordingly, their \eddr\ are over-estimated. If we exclude high-$\av$ sources (with $\av> 1$ mag), we only have five AGNs exceeding $\eddr > -0.5$. 
(Mrk~359, Mrk~382, Mrk~783, 2MASXJ08551746-2854218 and 2MASXJ21344509-2725557). 
After applying the BLR extinction correction, the extinction-corrected Eddington ratio estimates are in the range $-2.43 < \eddrc < -0.37$, with a median $\eddrc = -1.16$. 

In the right panel of Figure \ref{fig:kohovcaglar}, we plot \delm\ versus the \eddr, including both extinction-corrected and uncorrected sets of estimates (affecting both axes). 
The BLR extinction correction causes the $\mbh$ estimates to increase, and the \eddr\ estimates to accordingly decrease, by $\sim$0.1 dex (on average). However, for AGNs with high levels of BLR extinction ($\av > 2$ mag), the raw \eddr\ can be over-estimated by $\sim$0.4 dex. For the most extreme case in our sample, NGC 1365 (BAT ID 184; $\av = 4.1$ mag) the change in \eddr\ is 0.75 dex. 
To investigate trends in this parameter space, we first divide the data points into two bins in Eddington ratio, low and high, with equal numbers of data points in each. We see that the median \eddr\ values show a decreasing trend for both extinction corrected and uncorrected estimates. 
This trend is then also confirmed through a formal Spearman correlation test (for all extinction-corrected data points), which results in $\rho = -0.37\pm{0.05}$ ($p\ll0.01$). 

We now discuss the possibility of whether AGNs in the nearby universe are growing towards the $\mbh - \sig$ relation. To understand this, we first estimate the physical accretion rates of our AGNs, assuming $\Lbol = \eta \dot{M} c^2$ and a universal radiative efficiency $\eta=0.1$. 
The resulting accretion rates are in the range of $10^{-4} \leq \dot{M} \leq 1.4 \, M_{\odot}\,{\rm y}^{-1}$ with a median of $\dot{M} = 0.085\, M_{\odot}\,{\rm y}^{-1}$. Most of our Type 1 BASS AGNs are thus growing with low accretion rates and at sub-Eddington levels. Here, if we assume that the offset of our sources from the canonical $\mbh - \sig$ relation is explained by the ongoing growth of SMBHs destined to ``catch up'' with their host galaxies, we can actually estimate the required duration of the active accretion (i.e., AGN) phase for achieving this.
For our BASS-based sample, the median $\mbh$ is $\sim 10^{7.5}\,\Msun$ and the average \delm\ is 0.3 dex, which for the scenario we consider here would imply the SMBHs have to grow by a factor of $\sim2$, or by $\sim 10^{7.5}\,\Msun$ in mass. Given the aforementioned median accretion rate of $0.085\, M_{\odot}\,{\rm y}^{-1}$, this yields an SMBH growth time (AGN phase) of $\sim 10^{8.5}$ yr (i.e., $\sim0.4$ Gyr) is needed for eliminating the offset between BH mass estimates from $M_\mathrm{BH,BLR}$ and $M_\mathrm{BH,M-\sig}$. 
On one hand, this rough estimate for the AGN lifetime is consistent with that is implied from the integrated accretion density of distant AGNs by previous works (10$^{7-9}$ yr; i.e., the Soltan argument; \citealt{Soltan,Martini,Marconi2004}).
On the other hand, more recent evidence for the episodic nature of AGN accretion, with luminous episodes lasting as little as $\sim 10^{5}$ yr \cite[or even less; see, e.g.,][and references thereing]{Schawinski2015,Shen2021}, means that closing the BH mass gap through persistent growth at the observed (low) accretion rates is very unlikely. Of course, SMBHs are expected to undergo a wide range of accretion rates, from sub-Eddington to super-Eddington levels, during the AGN life cycle, therefore potentially expediting the process.

\subsection{Correlation Matrices of the Observable 
Parameters}\label{sec:corr_pca}

In the preceding sections, we have directly addressed several potential correlations between \delm, which by itself is derived from the AGN luminosity, broad line width, and \sigs\, and key AGN properties. 
Our analysis revealed some statistically significant trends and refuted others, while facing several observational biases. This motivates us to assess more systematically which basic observables and derived quantities are correlated with each other.

To this end, we compute the correlation matrix for the following quantities: \sig\, $\log(1+z)$, $FWHM_{\mathrm H\beta}$, $L_\mathrm{H\beta}$, $M_\mathrm{BH,H\beta}$, $FWHM_{\mathrm H\alpha}$, $L_\mathrm{H\alpha}$, $M_\mathrm{BH,H\alpha}$, 
$\nh$, $\av$, $L_\mathrm{Bol}$, $\lambda_\mathrm{Edd}$, and \delm. Here, we note that all quantities mentioned in the previous sentence are used with their raw observed values and no extinction correction was applied to them. In Table \ref{tab_spearman} (in Appendix \ref{sec_spearman_rank_order}), we present the correlation matrix computed using the Spearman rank-order correlation test. 
We use three different colors to indicate the significance of the correlation (or lack thereof): black and blue numbers represent significant correlations, with $p\ll0.01$ and $<0.05$, respectively, while red numbers represent null results (i.e. lack of a significant correlation), with $p\gg0.1$. For example, both the $FWHM_\mathrm{H\alpha}$ and $M_\mathrm{BH,H\alpha}$ pair of parameters, or the $L_\mathrm{H\beta}$ and $L_\mathrm{H\alpha}$ pair, show strong correlations ($\rho = 0.79\pm{0.02}$ and $0.85\pm{0.01}$, respectively; both with $p\ll0.01$). These example results are not surprising given that the parameters in both comparisons are, by definition, closely interlinked. 

The BLR extinction ($\av$) shows a statistically significant ($p\ll0.01$) \emph{anti}-correlation with $L_\mathrm{H\beta}$,  $L_\mathrm{5100}$, $L_\mathrm{H\alpha}$, and \delm\ (with $\rho=$ -0.49$\pm{0.03}$, -0.43$\pm{0.03}$, -0.50$\pm{0.03}$ and -0.38$\pm{0.07}$, respectively), while on the other hand showing a significant correlation with \eddr\ ($\rho = 0.42\pm{0.05}$, $p\ll0.01$). 
We also see that many properties are correlated with redshift, including  $L_\mathrm{H\beta}$, $L_\mathrm{5100}$, $L_\mathrm{H\alpha}$, $L_\mathrm{Bol}$ and \delm\ ($\rho$ = 0.51$\pm{0.03}$, 0.54$\pm{0.03}$, 0.66$\pm{0.02}$, 0.83$\pm{0.05}$ and 0.33$\pm{0.03}$ respectively; all with $p\ll0.01$). As discussed above, the correlations between redshift and the various luminosities are a manifestation of the flux-limited nature of our parent sample from the survey. 
We can finally also see the anti-correlation between \delm\ and both $\lambda_\mathrm{Edd}$ and $\av$, which was also discussed above. In Appendix \ref{sec:PCA}, we present the principal component analysis results in order to identify the main parameters driving the variance in our data sets (see Table \ref{tab:PCA}).

\section{Conclusion}\label{sec:summary}

We presented a study of stellar velocity dispersions (\sigs) in the host galaxies of a large sample of broad-line (Type 1), ultra-hard X-ray selected low-redshift AGNs. 
Our $z \leq 0.08$ AGNs are drawn from the flux-limited 105-month Swift-BAT catalog, and our analysis relies on optical spectroscopy obtained as part of the BASS project. 
We provide new measurements of \sigs, obtained for both the \cat and the \cahk+\mgi\ spectral complexes, for a total number of 173 AGNs. This work is one of the largest \sig\ investigations for Type 1 AGNs. Using the broad emission line measurements and derived $\mbh$ estimates made available through BASS/DR2 \citep{Mejia_Broadlines}, we compare our results with the established $\mbh - \sig$ relations. 
Our main findings are as follows: 

\begin{itemize}

\item  The average offset between \sigb\ and \sigr\ measurements is essentially negligible, at $0.002\pm0.001$ dex, and this shows that these two distinct spectral regimes provide highly consistent \sig\ measurements. 

\item We fit new $\mbh - \sigs$ relations using various data sets based on our sample and measurements. 
The slopes we find are significantly shallower than those reported in the literature for inactive galaxies. 
This result agrees with, and strengthens, the conclusion of previous studies of broad-line, low-redshift AGNs. Using an appropriate, AGN-based $\mbh - \sig$ relation for SE prescriptions may thus be advisable.

\item We show that BLR extinction plays an important role in single-epoch (SE, or virial) \mbh\ estimates, in that it causes the underestimation of \mbh\ and---consequentially---the overestimation of the Eddington ratios (\eddr).

\item We have looked into differences between SE and \sigs-based \mbh\ estimates, \delm, where the latter are based on the canonical $\mbh-\sigs$ relation of KH13. 
We found that \delm\ shows statistically significant correlations with both redshift and luminosity, however, these trends are likely driven by the nature of the survey and are mutually degenerate.

\item After applying the extinction correction to $\mbh$ measurements, we find Eddington ratios in a range of $-2.43 < \eddrc < -0.37$. In addition, the resulting physical accretion rates (ranging  $10^{-4} \lesssim \dot{M} \lesssim 1.40$ $M_{\odot}$ $yr^{-1}$) suggest that our broad-line BASS AGNs are growing at sub-Eddington levels.

\end{itemize}

The implications of our analysis are not yet fully understood, and further research is necessary to gain a clearer understanding of all biases and discrepancies between AGN and inactive galaxy samples. We specifically foresee further observations with high-resolution instruments, aboard the Hubble Space Telescope and/or James Webb Space Telescope, to directly probe how different types of host galaxy morphology might affect our interpretation of the $\mbh - \sigs$ relations for powerful AGNs. Additionally, more black hole mass measurements using near-infrared spectroscopy could help reduce the effects caused by dust, which can significantly interfere with our current understanding of AGN populations and of their relationship with their host properties (as we have demonstrated here). The results presented in this work thus aim to serve as a reference point for forthcoming, more detailed studies of the $\mbh - \sigs$ relation.

While the large size and high completeness of the sample used for our analysis present significant progress in studying the $\mbh-\sigs$ relation, and extinction effects, in low-redshift AGNs, it also highlights areas where more progress in terms of the census of low-redshift AGNs and their hosts is direly needed. Specifically, new \& upcoming X-ray missions, such as extended Roentgen Survey with an Imaging Telescope Array on the Spectrum-Roentgen-Gamma \citep[][]{Predehl2021}, Advanced Telescope for High Energy Astrophysics \citep{AthenaWhite}, and the Advanced X-ray Imaging Satellite \citep[][]{2018Mushotzky}, will greatly improve our ability to construct yet larger, more complete AGN samples at high redshift for which homogeneous \& robust spectral analysis can be obtained, to deduce key properties both in the X-ray and also in optical regimes thanks to spectroscopic surveys such as Sloan Digital Sky Survey-V \citep[][]{2017Kollmeier,Almeida2023} or the 4-metre Multi-Object Spectroscopic Telescope \citep[][]{2016deJong}. This will further help to break any outstanding degeneracies between AGN luminosities, BH masses, accretion rates and states, and host galaxy types and properties.

\section*{Acknowledgments}
We acknowledge support from NASA through ADAP award NNH16CT03C (M.K.); 
the Israel Science Foundation through grant No. 1849/19 (B.T.); 
the European Research Council (ERC) under the European Union's Horizon 2020 research and innovation program, through grant agreement No. 950533 (B.T.);
FONDECYT Regular 1230345 (C.R), 1190818 (F.E.B.) and 1200495 (F.E.B); ANID grants CATA-Basal AFB-170002 (F.E.B.), ACE210002 (F.E.B.) and FB210003 (C.R., F.E.B.); Millennium Science Initiative Program  – ICN12\_009 (F.E.B.); Fondecyt Iniciacion grant 11190831 (C.R.); the Korea Astronomy and Space Science Institute under the R\&D program (Project No. 2023-1-868-03) supervised by the Ministry of Science and ICT, the National Research Foundation of Korea grant NRF-2020R1C1C1005462, and the Japan Society for the Promotion of Science ID: 17321 (K.O.); Fundaci\'on Jes\'us Serra and the Instituto de Astrof{\'{i}}sica de Canarias under the Visiting Researcher Programme 2023-2025 agreed between both institutions.  ACIISI, Consejer{\'{i}}a de Econom{\'{i}}a, Conocimiento y Empleo del Gobierno de Canarias and the European Regional Development Fund (ERDF) under grant with reference ProID2021010079, and the support through the RAVET project by the grant PID2019-107427GB-C32 from the Spanish Ministry of Science, Innovation and Universities MCIU. This work has also been supported through the IAC project TRACES, which is partially supported through the state budget and the regional budget of the Consejer{\'{i}}a de Econom{\'{i}}a, Industria, Comercio y Conocimiento of the Canary Islands Autonomous Community.Conselho Nacional de Desenvolvimento Cient\'{i}fico e Tecnol\'ogico  ( CNPq, Proj. 311223/2020-6,  304927/2017-1 and  400352/2016-8), Funda\c{c}\~ao de amparo 'a pesquisa do Rio Grande do Sul (FAPERGS, Proj. 16/2551-0000251-7 and 19/1750-2), Coordena\c{c}\~ao de Aperfei\c{c}oamento de Pessoal de N\'{i}vel Superior (CAPES, Proj. 0001, R.R.); NASA ADAP Grant Number 80NSSC23K0557 (T.T.A.). L.A.S. acknowledges financial support from the Swiss National Science Foundation (SNSF). This work was performed in part at Aspen Center for Physics, which is supported by National Science Foundation grant PHY-1607611. The work of DS was carried out at the Jet Propulsion Laboratory, California Institute of Technology, under a contract with NASA.

We acknowledge that the data, which is used in this work, is obtained through the following observatories: the European Organisation for Astronomical Research in the Southern Hemisphere (ESO), the Palomar Observatory, the Southern Astrophysical Research (SOAR), the W.M. Keck Observatory, and the various stages of the Sloan Digital Sky Survey (SDSS). A full list of observing facilities and program numbers can be found in the main DR2 Catalog paper \cite[][see the acknowledgments there]{Koss_DR2_catalog}.

\bibliography{main.bbl}{}
\bibliographystyle{aasjournal}

\begin{appendix}

\section{\lowercase{p}PXF fit results}\label{sec:sigb_sigr_plots}

\begin{figure}
\centering
\includegraphics[width=0.475\textwidth]{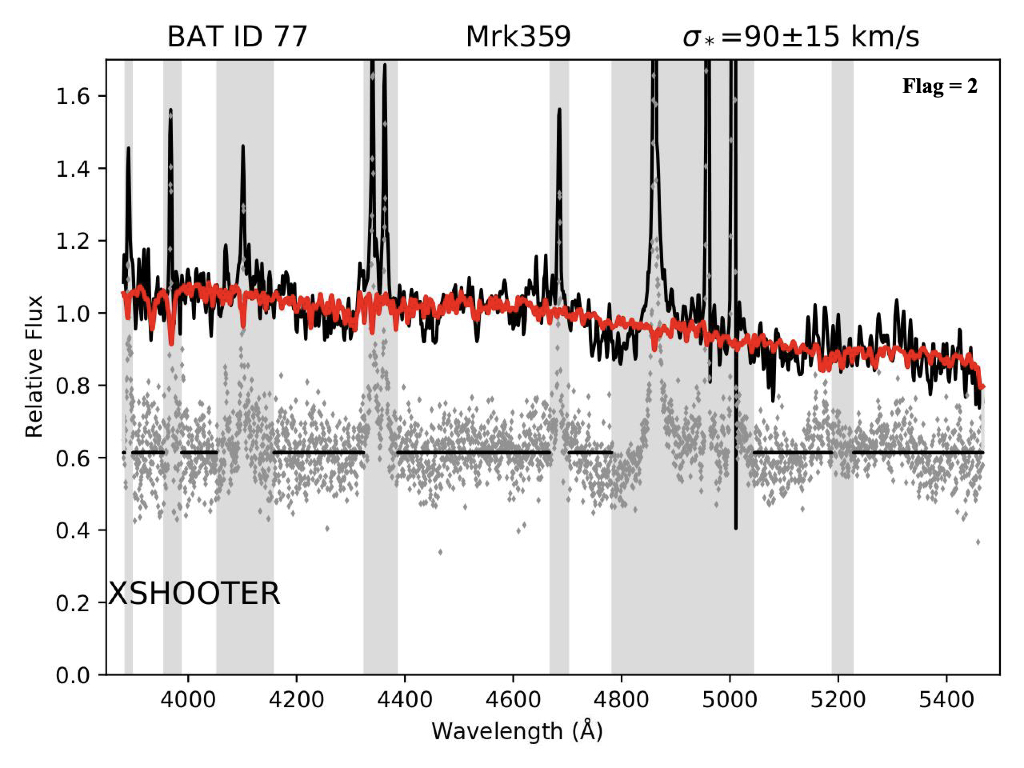}
\includegraphics[width=0.475\textwidth]{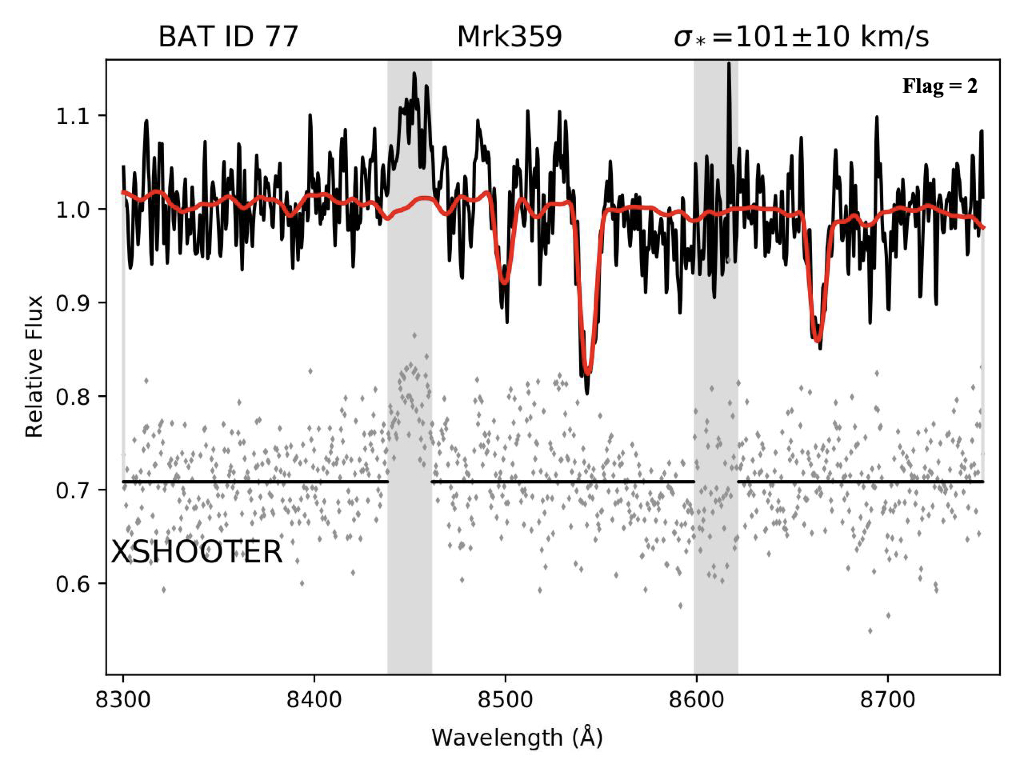}
\includegraphics[width=0.475\textwidth]{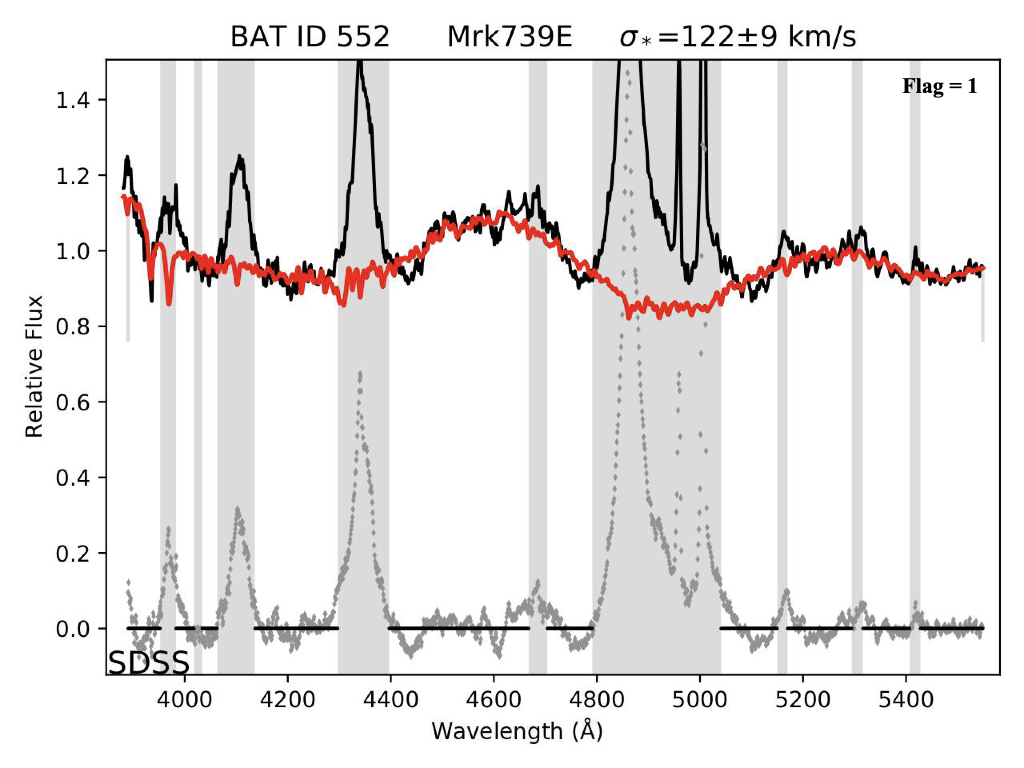}
\includegraphics[width=0.475\textwidth]{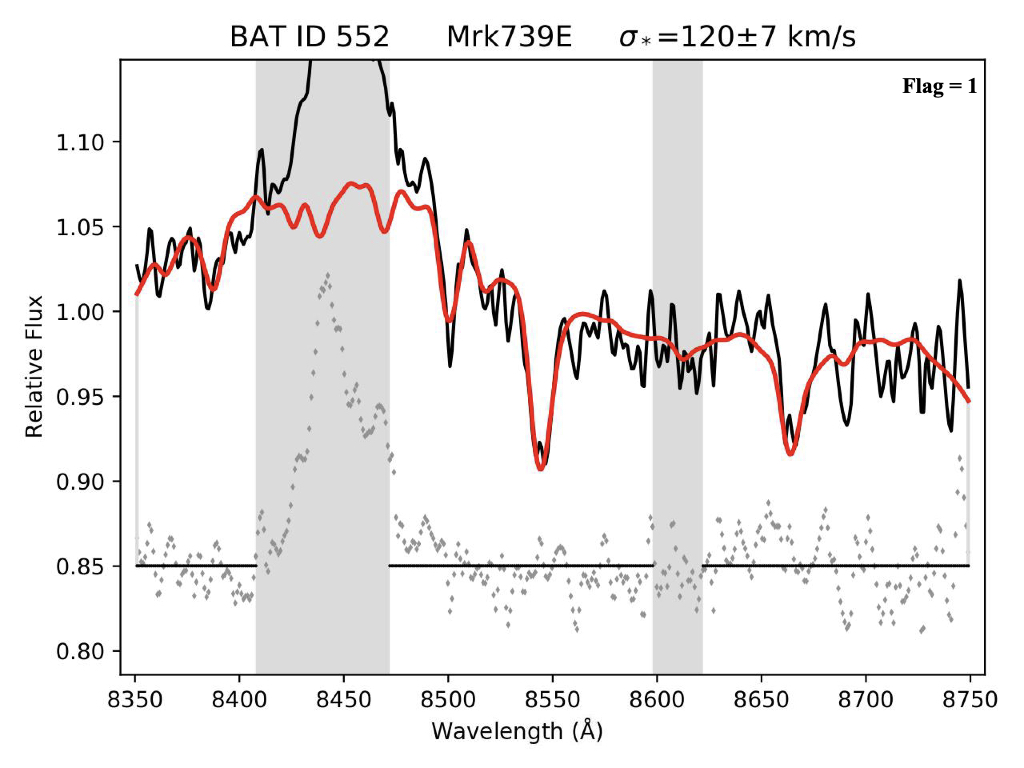}
\includegraphics[width=0.475\textwidth]{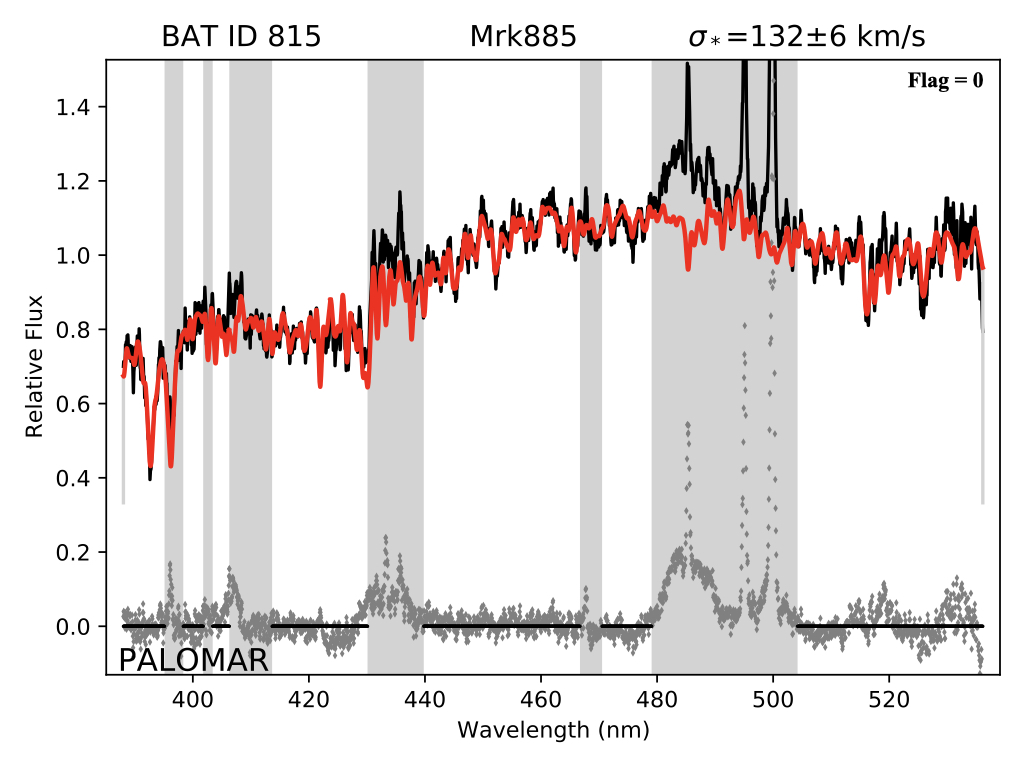}
\includegraphics[width=0.475\textwidth]{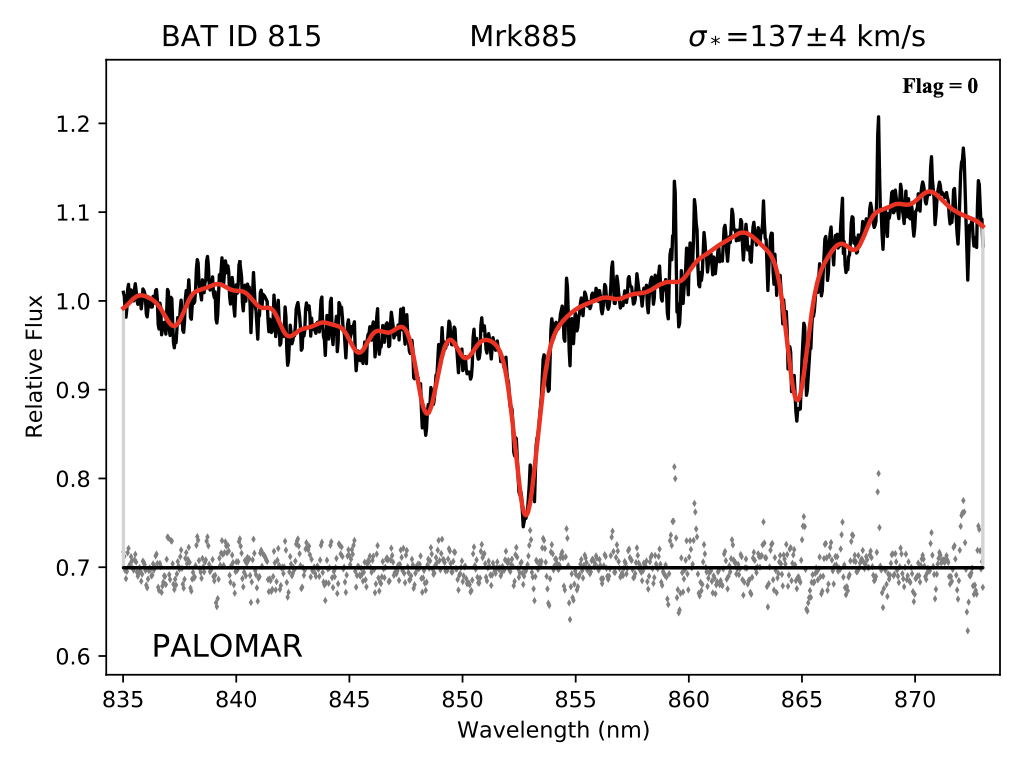}
\caption{Examples of the successful \sigb\ (\textbf{left panels}) and \sigr\ (\textbf{right panels}) fitting plots for SDSS, VLT/X-shooter, and Palomar/Double Spectrograph data (from top to bottom). The complete figure set (529 images) is available in the online journal.}
\label{fig:sigb_sigr_plots}
\end{figure}

\begin{figure}
\centering
\includegraphics[width=0.475\textwidth]{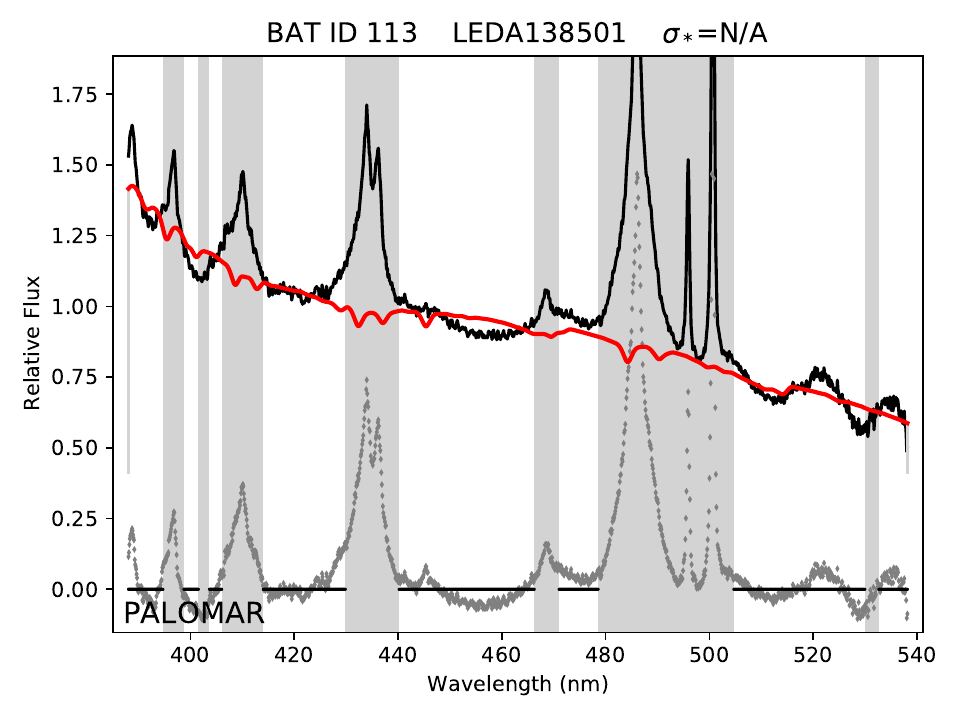}
\includegraphics[width=0.475\textwidth]{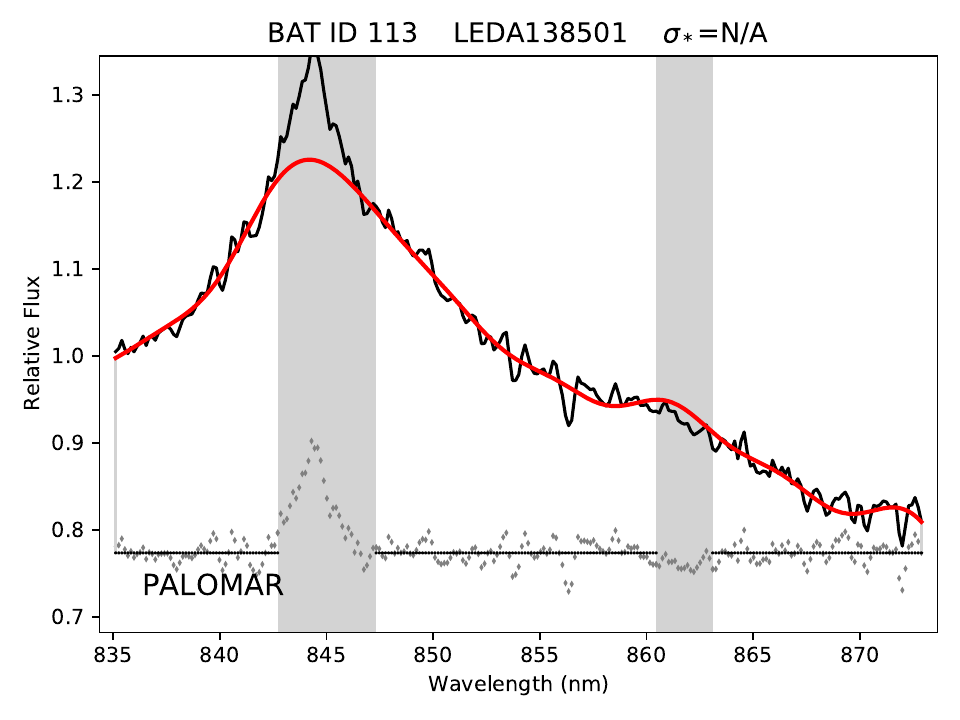}
\includegraphics[width=0.475\textwidth]{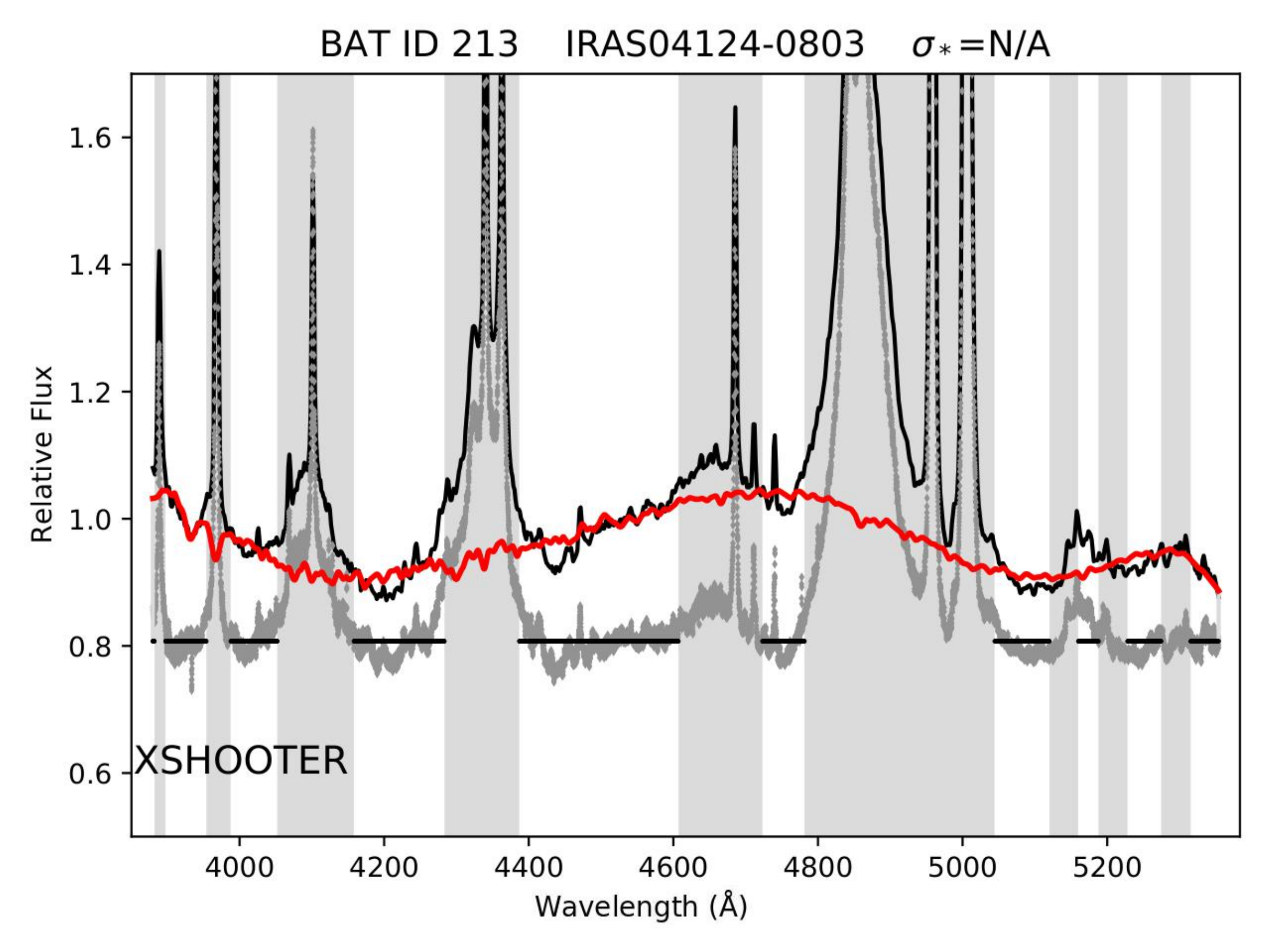}
\includegraphics[width=0.475\textwidth]{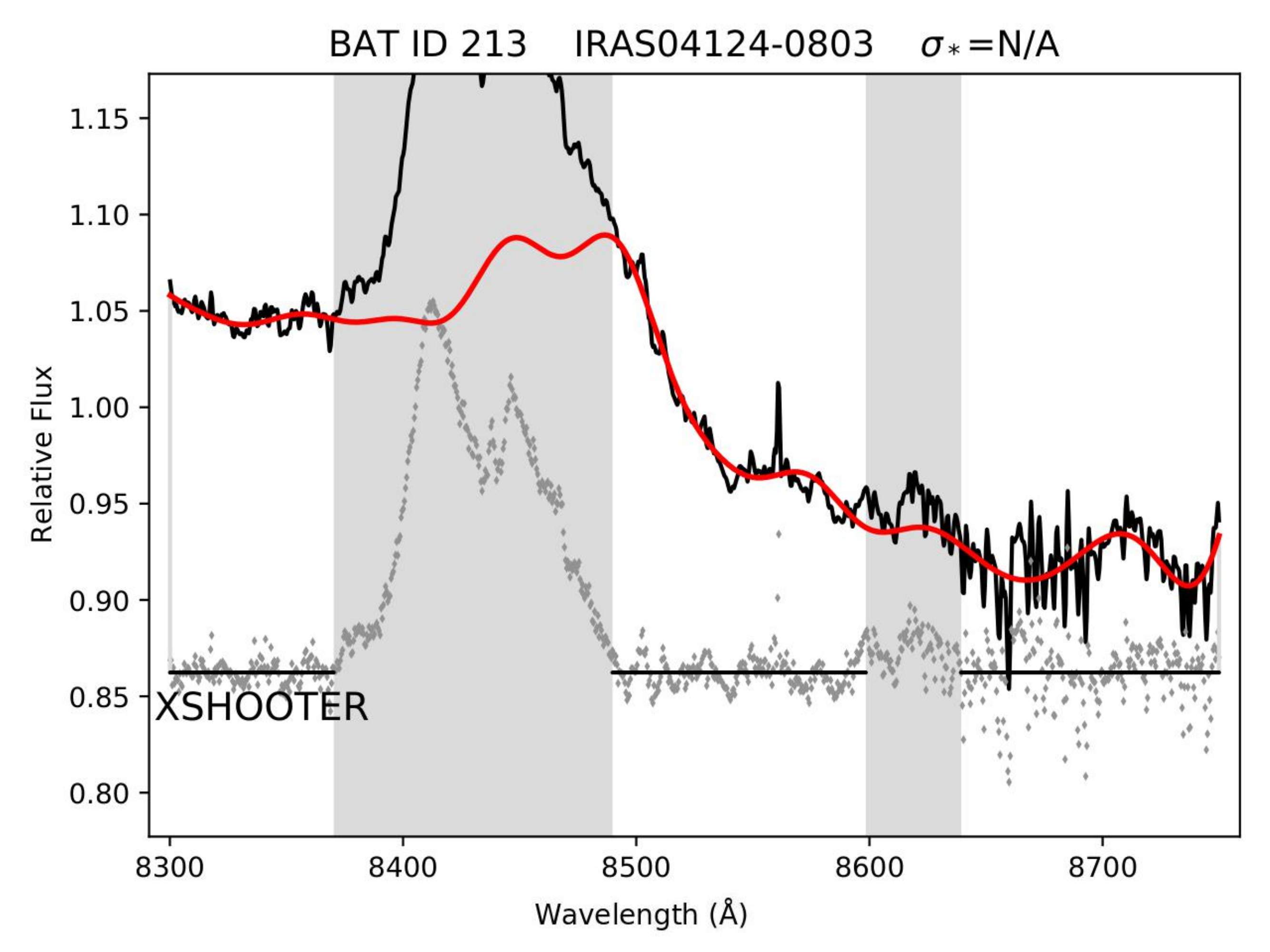}
\includegraphics[width=0.475\textwidth]{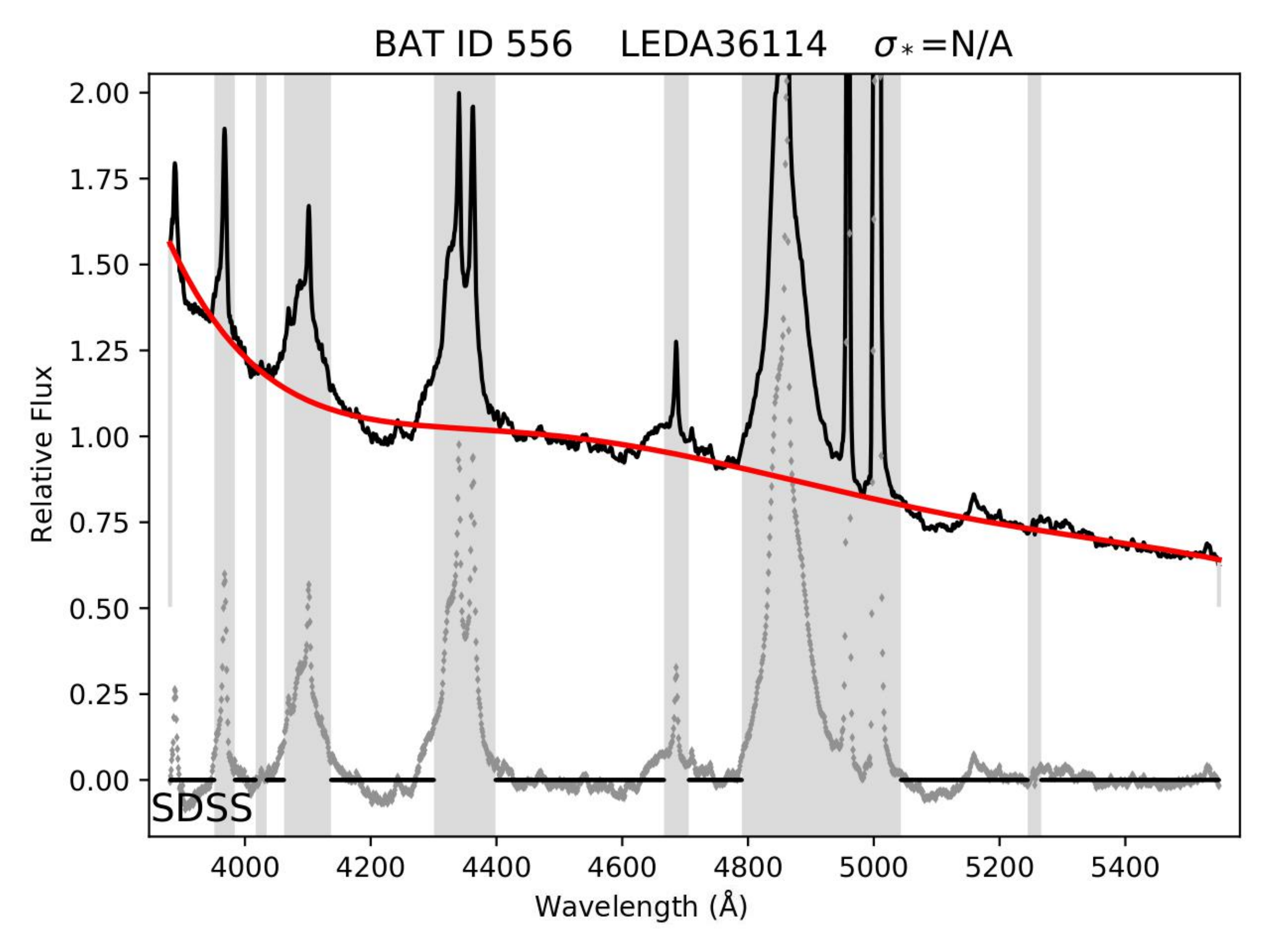}
\includegraphics[width=0.475\textwidth]{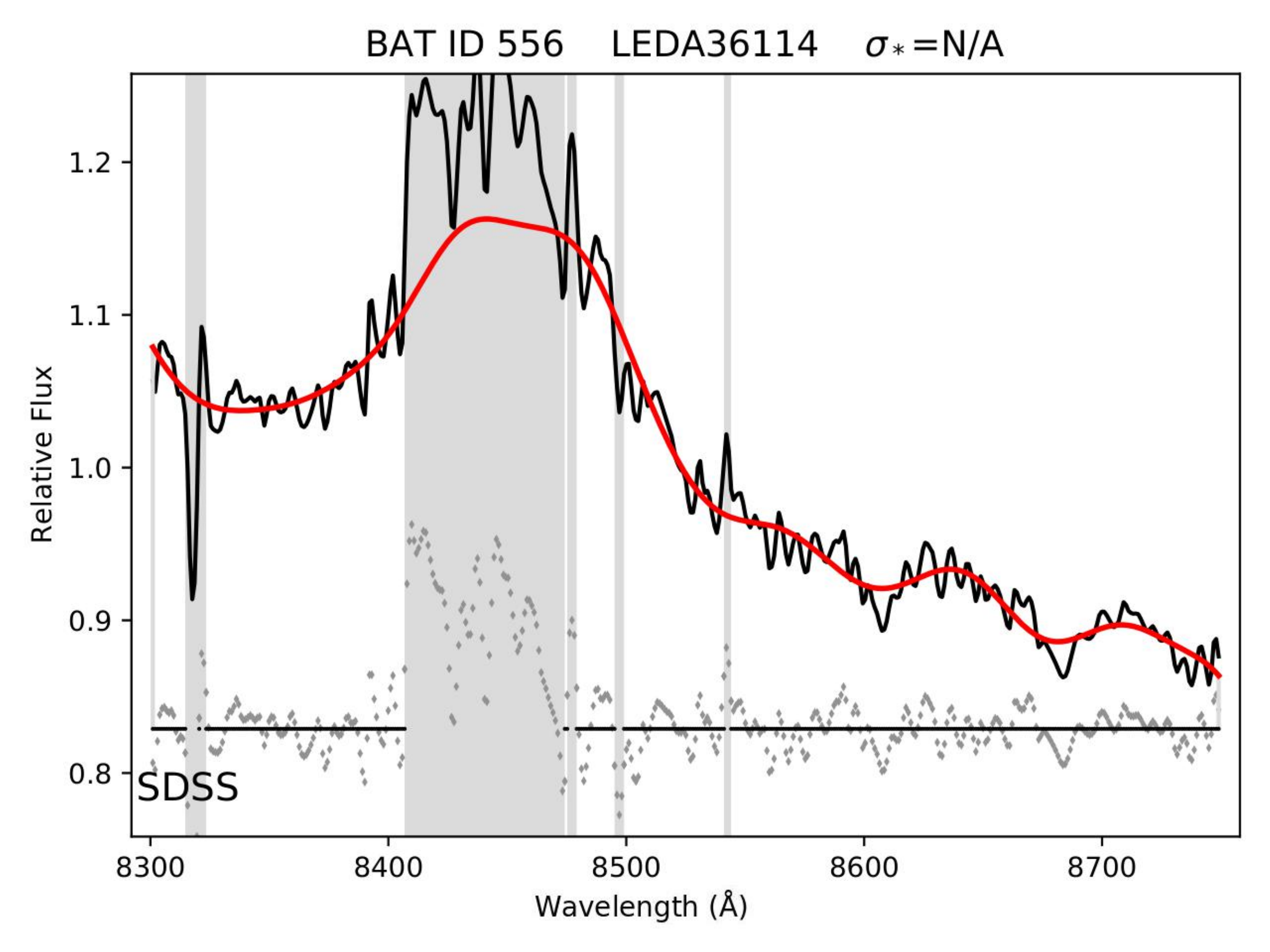}
\caption{Examples of the failed \sigb\ (\textbf{left panels}) and \sigr\ (\textbf{right panels}) fitting plots for Palomar/Double Spectrograph, VLT/X-shooter and SDSS data (from top to bottom). The complete figure set (529 images) is available in the online journal.}
\label{fig:sigb_sigr_plots_fail}
\end{figure}

In Figure \ref{fig:sigb_sigr_plots}, we present three examples of successful spectral fits, yielding robust measurements of \sigb\ (left column) and \sigr\ (right column). 
We also show three examples of failed spectral fits in Figure \ref{fig:sigb_sigr_plots_fail} (again, for both \sigb\ and \sigr\/). 
In both cases, the examples shown are representative of our spectral setups and fitting results. 

\section{Comparison of $\sigma_\mathrm{\star}$ measurements from different instruments in our survey}\label{sec:sigb_sigr_plots_other_instruments}

\begin{figure*}[htb!]
    \centering
    \includegraphics[width=0.494\textwidth]{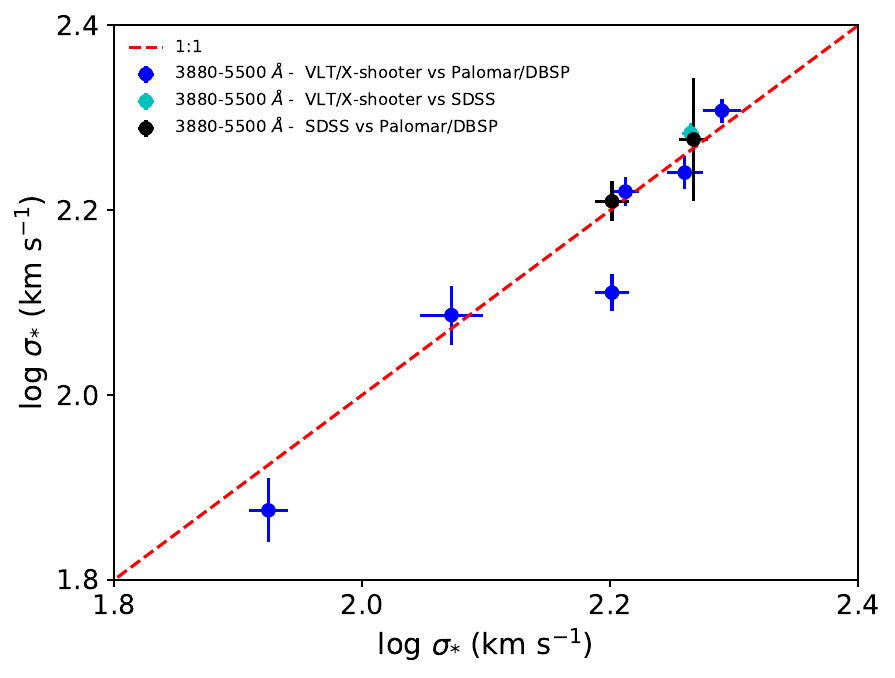}
    \includegraphics[width=0.494\textwidth]{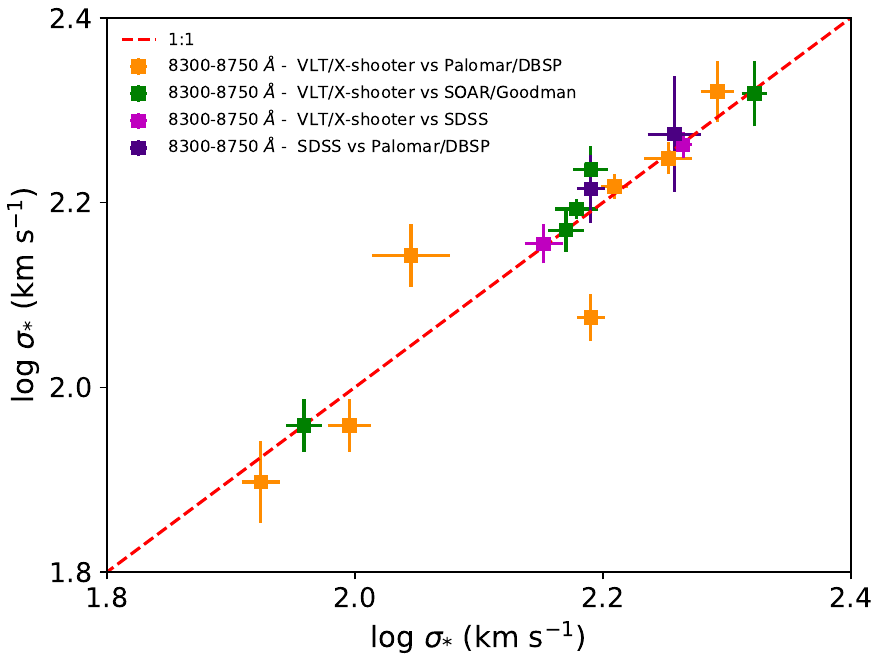}
    \caption{\textbf{Left:} The comparison of $\sigma_\mathrm{\star}$ measurements between the instruments used in this study for 3880-5500~\AA\ region. \textbf{Right} same as bottom left figure, but for 8350-8750~\AA\ region. The red dashed lines represent the 1:1 lines for visual aid.}
    \label{fig:all_sigma_instrument_comparison}
\end{figure*}

For a small subset of AGNs in our sample, where more than one optical spectrum is available, we were able to obtain (at least) two independent measurements of \sigs\ from the same spectral region.
In Figure \ref{fig:all_sigma_instrument_comparison}, we present a pair-wise comparison of these duplicate \sigs\ measurements.  
We find that our duplicate \sig\ measurements, obtained with different instruments, are highly consistent with each other, for both $\sigma_\mathrm{Blue}$ and \sigr\ measurements (i.e. both spectral regimes considered here). 
There is only one significant outlier from the 1:1 line in between the $\sigma_\mathrm{Blue}$ measurements and two significant outliers for \sigr\ measurements. 
The first outlier is BAT 197 (HE\,0351+0240) which shows $\approx$0.1 dex difference in \sigr\ measurements, whereas its $\sigma_\mathrm{Blue}$ measurements are essentially indistinguishable from each other (a difference of order 0.01 dex). 
The second outlier is BAT 562 (NGC\,3822) which shows an offset of 0.09 dex in $\sigma_\mathrm{Blue}$ and 0.11 dex in \sigr\ measurements. 
Such differences between \sig\ measurements may be caused by systematic uncertainties in some cases, including varying observational conditions, instrumental resolutions and/or aperture sizes, and the detailed spectral features of the templates used, in the spectral regions of choice. 
Such systematic uncertainties can be as large as the statistical uncertainties obtained from the pPXF resampling approach. A detailed explanation of the systematic uncertainties is given by \citet{Koss_DR2_sigs}. 
At any rate, Figure \ref{fig:all_sigma_instrument_comparison} demonstrates the robustness of our methodology and---given the potential uncertainties and caveats---is an encouraging result.

\section{Conversion between $\log (L^{\rm \lowercase{int}}_{\rm 14-150\,\kev}/\ergs)$ and $\log (L^{\rm \lowercase{obs}}_{\rm 14-195\,\kev}/\ergs)$}\label{sec:LXX}

\begin{figure}
\centering
\includegraphics[width=0.475\textwidth]{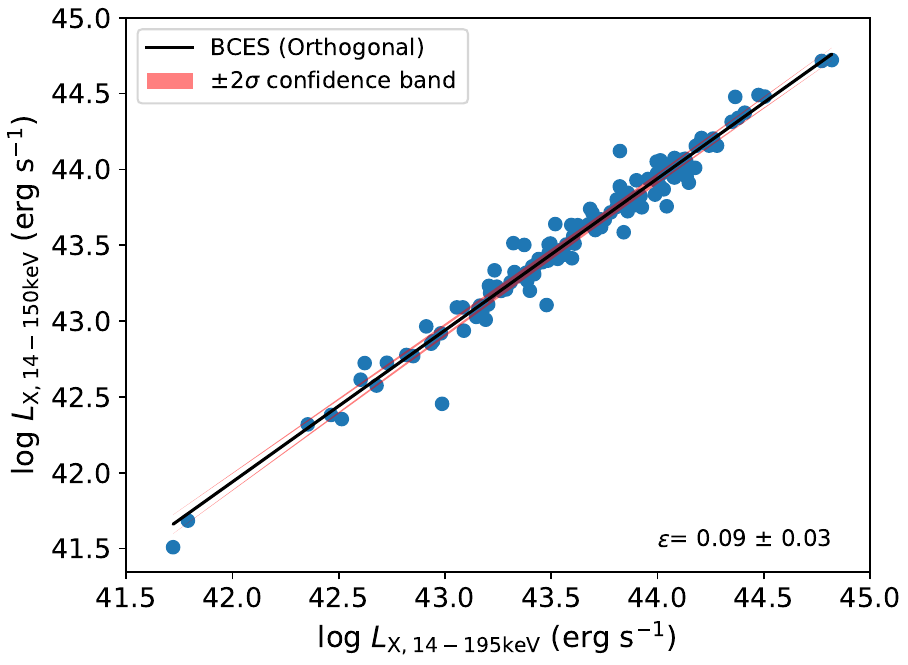}
\caption{The fitting results for the conversion between $\log (L^{\rm \lowercase{int}}_{\rm 14-150\,\kev}/\ergs)$ and $\log (L^{\rm \lowercase{obs}}_{\rm 14-195\,\kev}/\ergs)$.}
\label{fig:LXX}
\end{figure}

To measure $\log (L^{\rm int}_{\rm 14-150\,\kev}/\ergs)$ for our bonus sample of 55 AGNs, we first fit an orthogonal linear fit between $\log (L^{\rm int}_{\rm 14-150\,\kev}/\ergs)$ and $\log (L^{\rm obs}_{\rm 14-195\,\kev}/\ergs)$ for our sample of AGNs from the BAT 70-month catalog. Correspondingly, the resulting fit is found as follows: 

\begin{equation}
\log (L^{\rm int}_{\rm 14-150\,\kev}/10^{44}) = \log (L^{obs}_{\rm 14-195\,\kev}/10^{44}) - 0.06\pm{0.01} \: \: \: \: \ergs.
\end{equation}
The resulting intrinsic scatter ($\epsilon$) of 0.09$\pm{0.03}$ dex allows us to perform such a conversion confidently. Here, we note that we normalized the X-ray luminosities with the value of 10$^{44}$ and also fixed the slope to 1 in order to avoid the correlation between the slope and intercept. In Figure \ref{fig:LXX}, we present the conversation together with fitting results.

\section{Comparison of extinction-corrected and uncorrected black hole masses} \label{sec:comparison_extinction_correction_mbh}

Here, we note that the extinction in the BLR becomes somewhat important beyond $\av \simeq 1$. However, we stress that the majority of our sample (108 out of 165) have $\av < 1$. We also stress that the median difference between extinction-uncorrected and extinction-corrected \mbh\ is found to be 0.088 dex for our sample of AGNs. In only 32 cases, the difference exceeds 0.3 dex, and only six of them exceed 0.5 dex difference.

\section{The success rate of $\sigma_\mathrm{\star}$ measurements and various AGN properties}\label{sec:various_AGN_properties}

\begin{figure*}
\centering
\includegraphics[width=0.46
\textwidth]{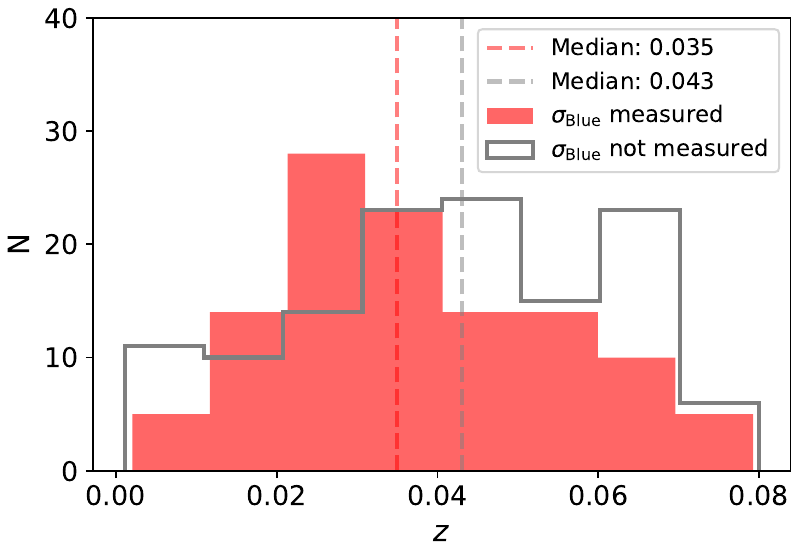}
\includegraphics[width=0.46\textwidth]{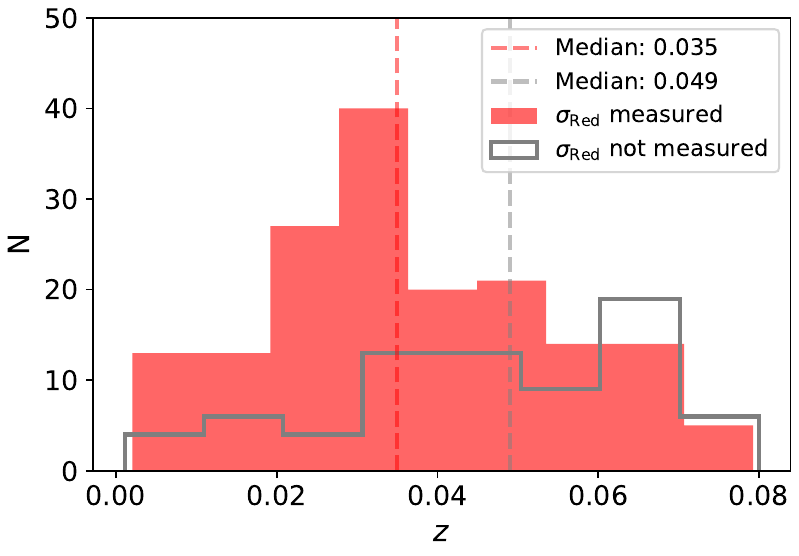}
\includegraphics[width=0.46\textwidth]{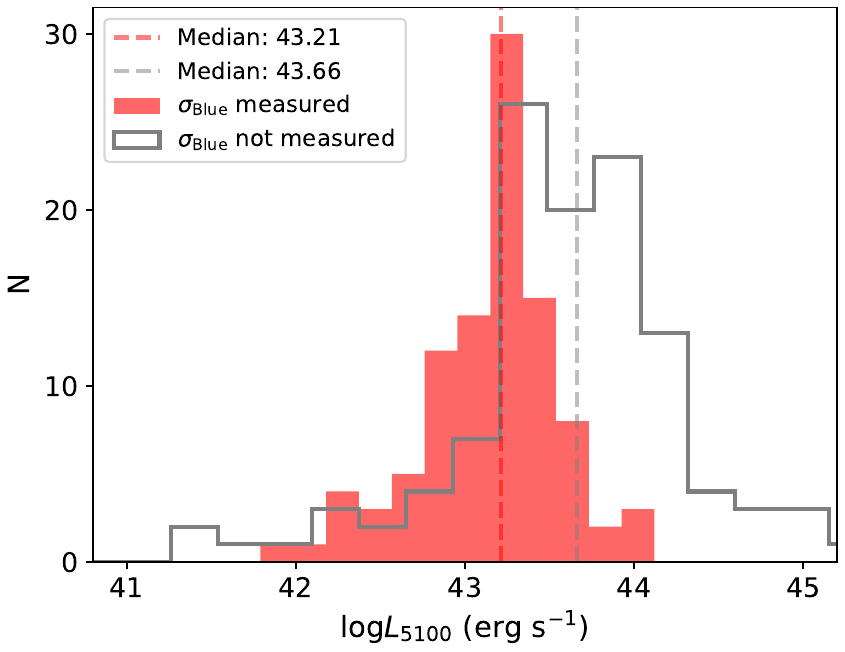}
\includegraphics[width=0.46\textwidth]{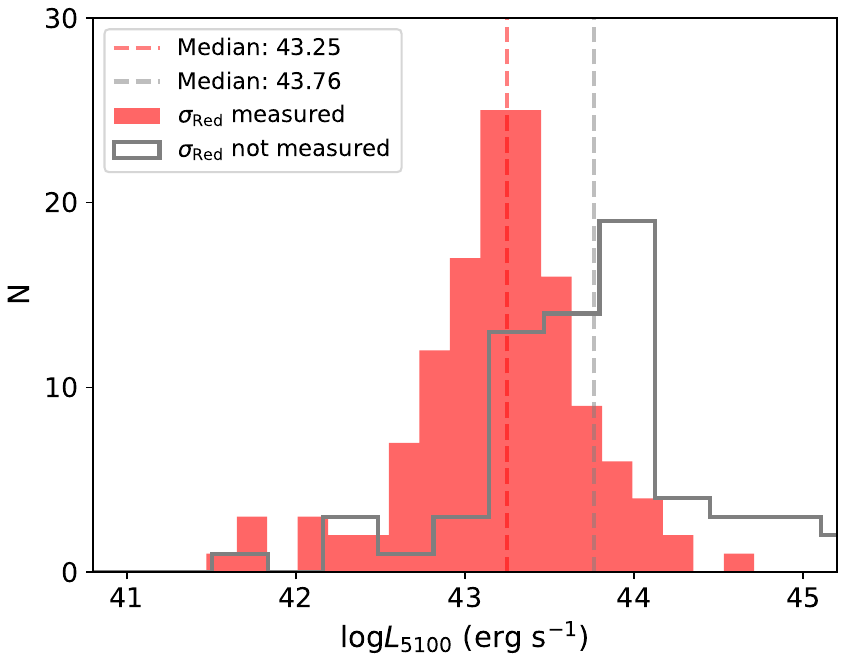}
\includegraphics[width=0.46\textwidth]{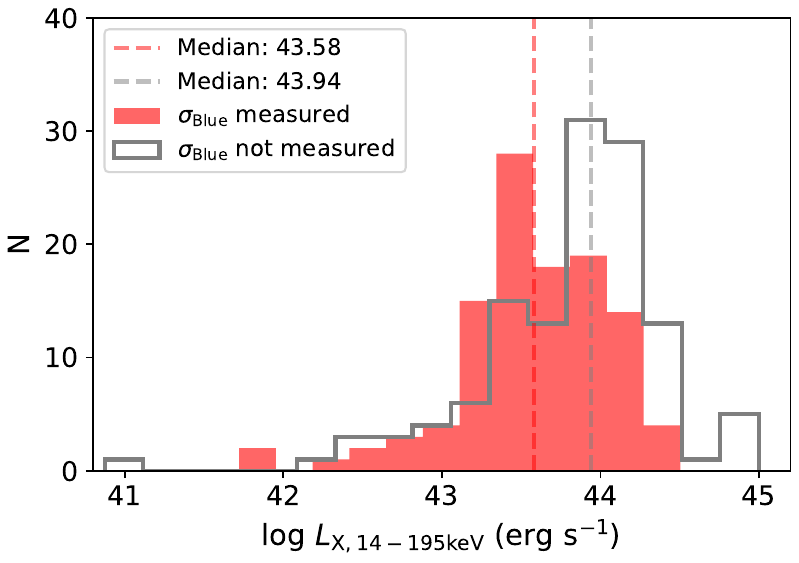}
\includegraphics[width=0.46\textwidth]{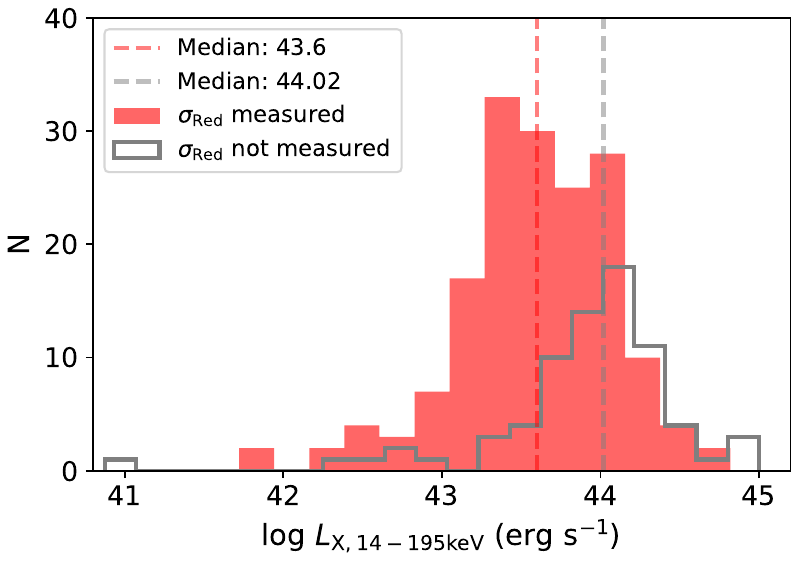}
\includegraphics[width=0.46\textwidth]{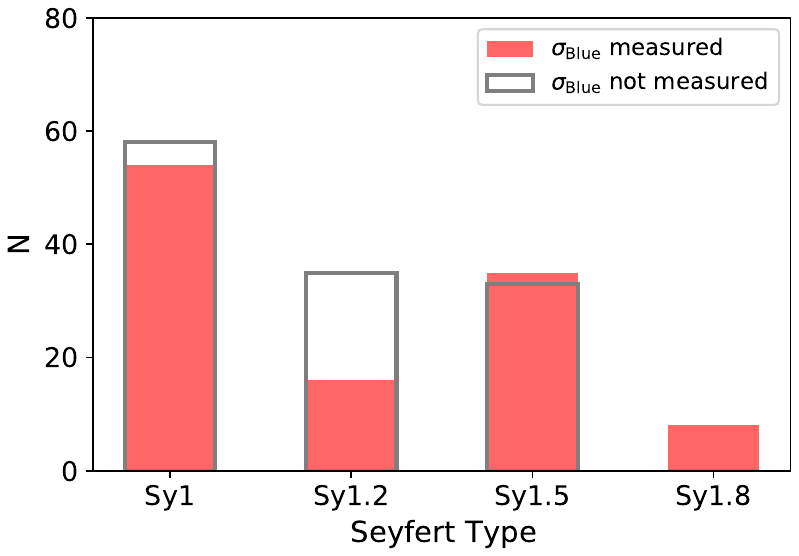}
\includegraphics[width=0.46\textwidth]{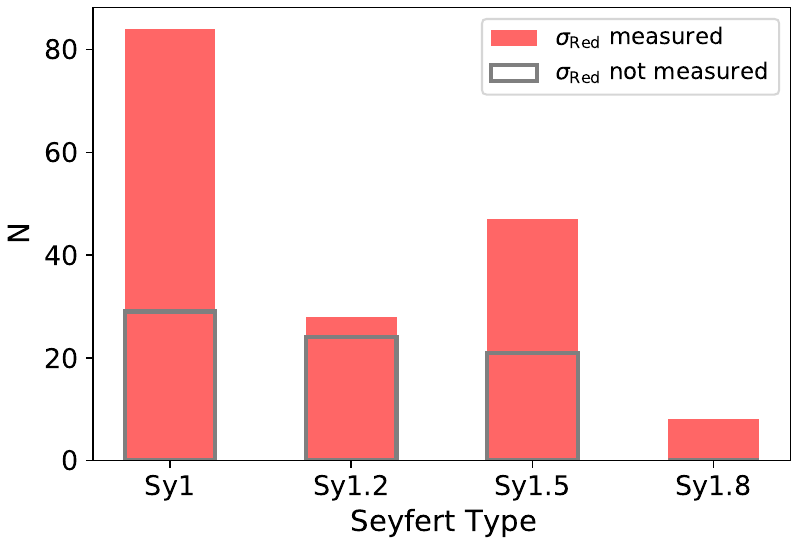}
\caption{The distributions of successful vs. failed \sig\ measurements with various AGN properties, for both \sigb\ (left column) and \sigr\ (right column).
From top to bottom, we show distributions of redshift $z$, optical continuum luminosity $L_{5100}$, ultra-hard X-ray luminosity $L_\mathrm{X, \: 14-195\: keV}$, and Seyfert sub-type.}
\label{fig_sig_d_nd_vsall}
\end{figure*}

In Figure \ref{fig_sig_d_nd_vsall}, we present the distributions of several key properties for our AGN sample, split into successful and failed \sigs\ measurements, and for both \sigb\ and \sigr\ measurements. We can see that the chance of obtaining successful \sig\ fittings decreases with increasing redshift (top panels) and/or with increasing luminosity (either optical or ultra-hard X-rays; second and third-row panels, respectively). 
The latter could be caused by either stronger continuum emission or broad line emission, both of which may dilute the stellar features. This might be considered a bias as the failed velocity dispersion objects tend to be the more luminous AGNs.
Finally, we see no significant link between our ability to measure \sigs\ and the Seyfert sub-types (bottom panels of Fig.~ \ref{fig_sig_d_nd_vsall}). 

\section{The $M_\mathrm{BH}$ - $\sigma_\mathrm{\star}$ relation for various data sets}\label{sec:M_sigma_plots_DS1234}

\begin{figure*}[hbt!]
\centering
\includegraphics[width=0.475\textwidth]{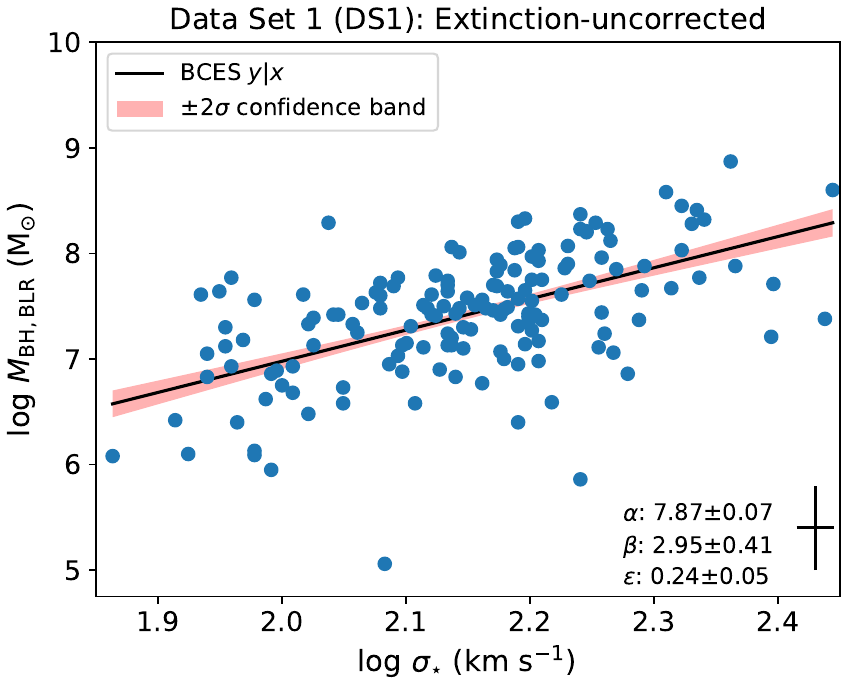}
\includegraphics[width=0.475\textwidth]{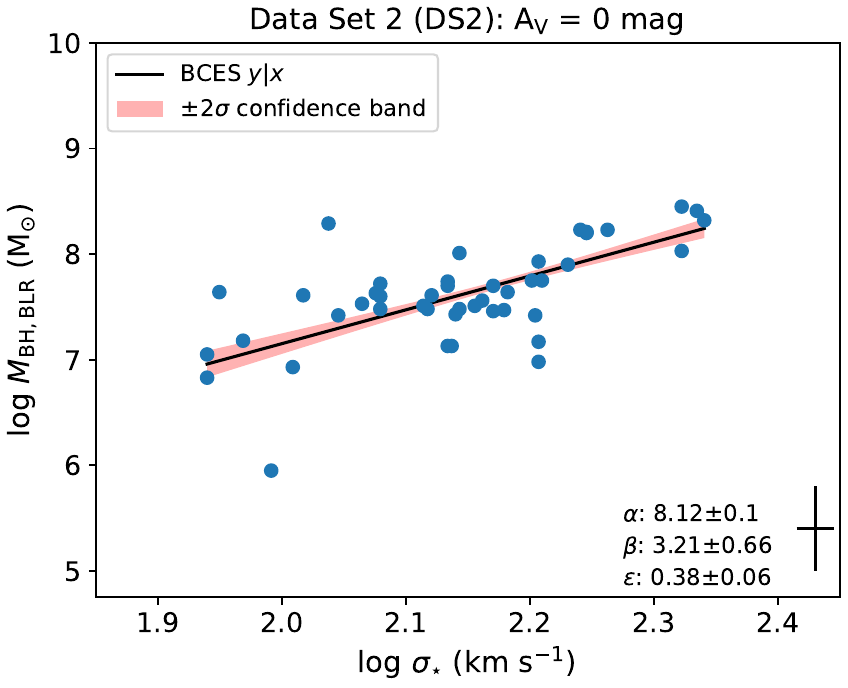}
\includegraphics[width=0.475\textwidth]{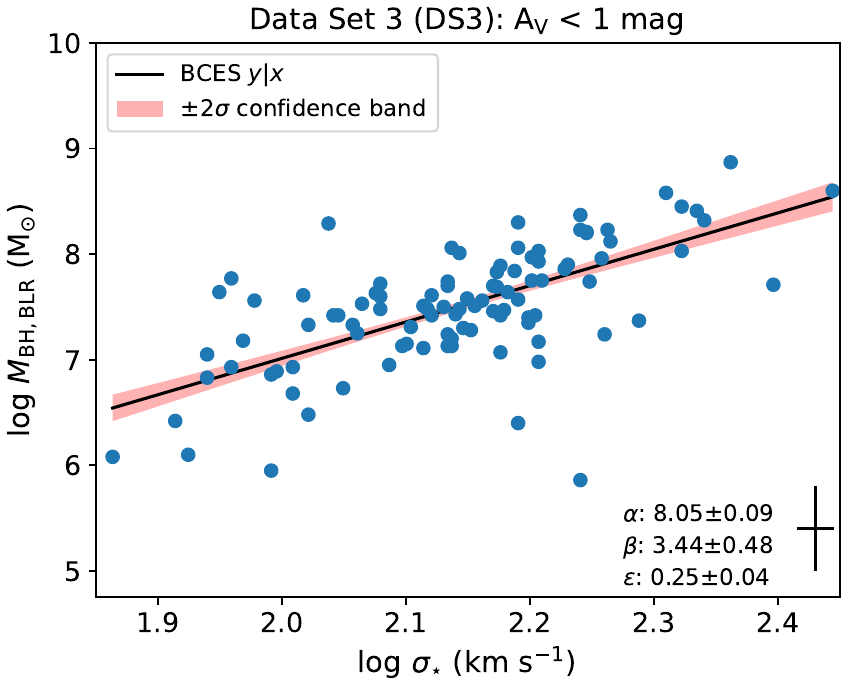}
\includegraphics[width=0.475\textwidth]{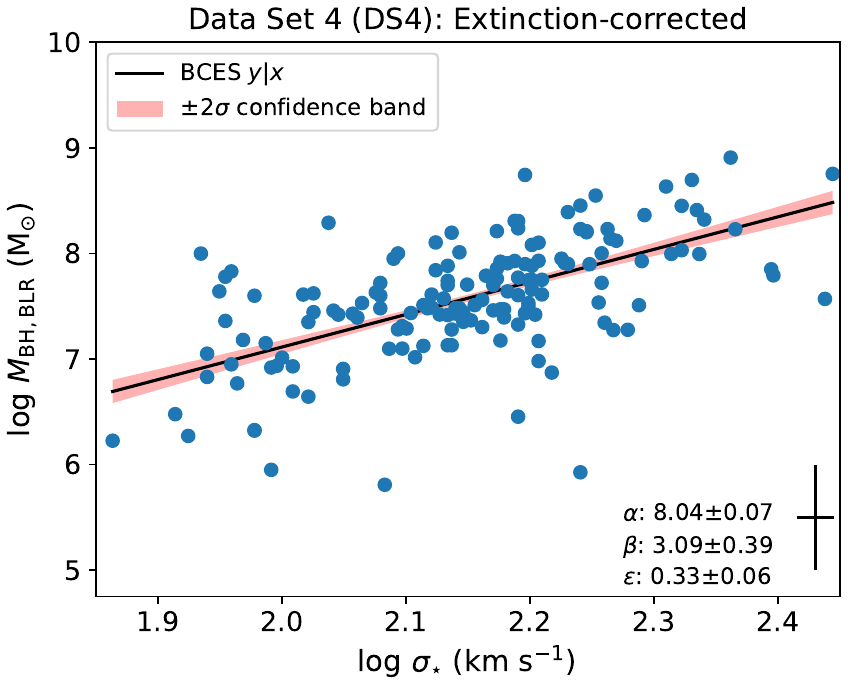}
\caption{The $\mbh - \sig$ relation of for different sub-samples in our sample: DS1 (\textbf{top left}), DS2 (\textbf{top right}), DS3 (\textbf{bottom right}), and DS4 (\textbf{bottom right}). The fitting parameters are as follows: $\alpha$ is the intercept, $\beta$ is the slope and $\epsilon$ is the intrinsic scatter of the $\mbh - \sig$ relation. We show the median uncertainty in \sigs\ and \mbh\ as a black plus sign for visual aid.}
\label{fig_m_sigma_for_DS}
\end{figure*}

In Figure \ref{fig_m_sigma_for_DS}, we present the four different data sets considered for our BASS-based $\mbh - \sig$ relations, and the corresponding best fits. 
The data sets are: 
all sources, with extinction-uncorrected measurements (DS1); 
sources with no signs of extinction ($\av= 0$; DS2);
sources with some, but not extreme, extinction ($\av< 1$; DS3); 
and all sources, but using extinction-corrected measurements (DS4). The resulting fits are shown with $\pm$2$\sigma$ confidence bands. Here, we remind the reader that NGC 7213 was excluded from our data sets due to the unreliable \mbh\ measurement. 

\section{Galaxy Morphology and Aperture Corrections}\label{sec_morp_aperture}

Various spectral observables and derived parameters for galaxy centers are known to depend on the aperture used during observations \cite[e.g.,][]{Jorgensen1995,Mehlert2003,Capp2006,FB2017}. This is particularly relevant for our work since galaxies generally have radial gradients in \sigs\ and in radial velocity. 
To investigate the importance of aperture effects and corrections for our analysis, we parameterize the \sigs\ gradient as a power law, that is:
\begin{equation}
\frac{\sigma_\mathrm{ap}}{\sigma_\mathrm{e}} =  \left( \frac{r_\mathrm{ap}}{r_\mathrm{e}} \right)^{\alpha} \, ,
\end{equation}
where $\alpha$ is the slope of the gradient, $r_\mathrm{e}$ is the effective radius, $r_\mathrm{ap}$ is the aperture radius adopted in our spectroscopic observations, $\sigma_\mathrm{ap}$ is the \sig\ we measure from these observations (at $r_\mathrm{ap}$) and $\sigma_\mathrm{e}$ is the stellar velocity dispersion at $r_\mathrm{e}$. 

A further complication arises from the finding that the slopes of such \sigs\ gradients depend on the galaxy morphologies \citep[and stellar masses for spiral galaxies,][]{FB2017}. 
To incorporate this dependence, we collected the morphological classifications available for our sample galaxies from the literature \citep{RC3,Vitores,Leda,Deo,Nair,delapp,Zoo}, and divide them into two categories, essentially split into early-type (E-like) and late-type (S-like) objects. 
Considering the average stellar masses of BAT AGN hosts ($\log(M_\mathrm{\star}/\Msun) = 10.28 \pm0.4$; \citet{2011Koss}), we adopt $\alpha = -0.055$ for early-type (E+L) galaxies and $\alpha=0.077$ for late-type (S), following \citep{FB2017}. 
Finally, we collected $K$-band effective radii from the Two Micron All Sky Survey \citep{2MASS} to compute the correction factor mentioned above.

\begin{figure*}[htb!]
    \centering
    \includegraphics[width=0.49\textwidth]{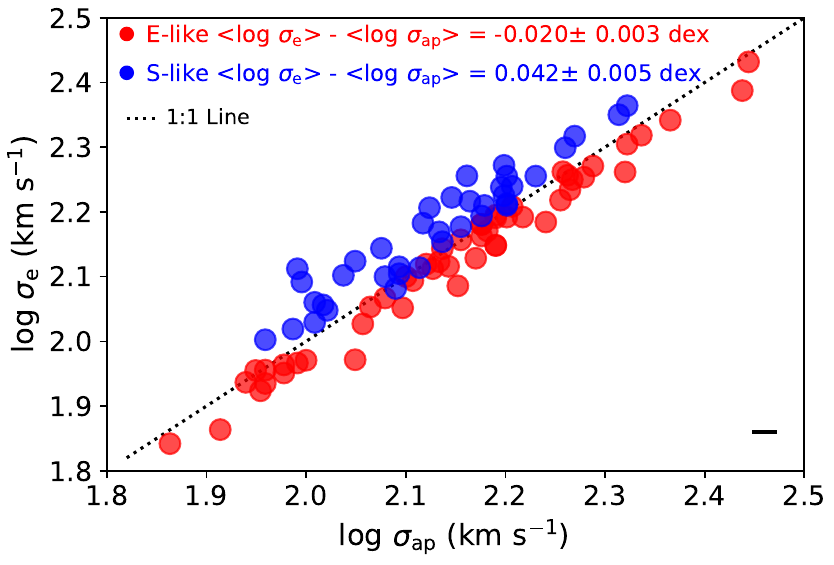}
    \includegraphics[width=0.49\textwidth]{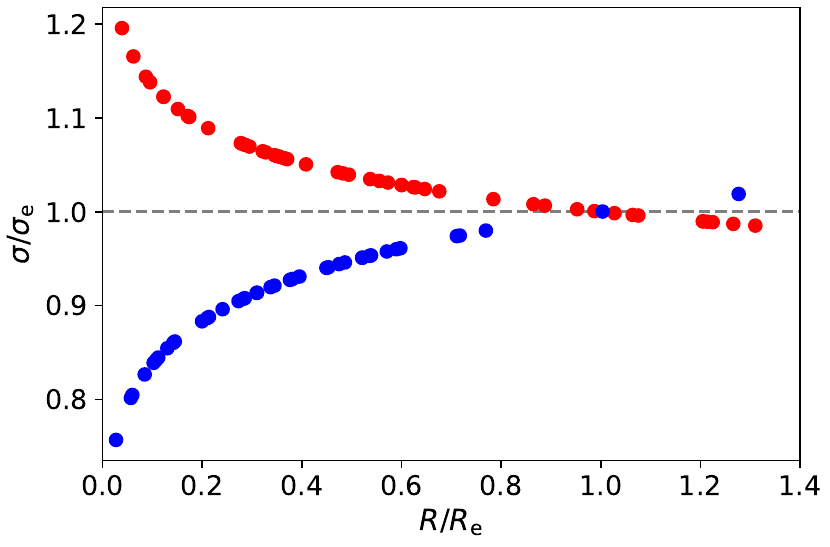}
    \caption{\textbf{Left:} The comparison between the \sig\ measurements obtained from our instruments' aperture sizes and effective radii. The black horizontal line represents the median uncertainty in \sigs\/, whereas the black dotted line represents 1:1 line. \textbf{Right:} The normalized \sigs\ profiles integrated within elliptical apertures with increasing semi-major radius.}
    \label{fig:aperture_all_sigma_instrument}
\end{figure*}

In Figure \ref{fig:aperture_all_sigma_instrument}, we demonstrate a direct comparison between the \sig\ measurements obtained from our observations (i.e., instrumental apertures) and those expected at the effective radii. 
We find an average offset of $-0.020\pm0.003$ dex between $\sigma_\mathrm{e}$ and $\sigma_\mathrm{ap}$ for E-like galaxies, and $0.042\pm0.005$ dex for S-like galaxies. 
The median relative error for our \sigs\ measurements ($\approx$0.02 dex; see Section~\ref{sec:sample_comp}) is thus comparable to the offset caused by aperture effects.
Aperture effects are thus unlikely to lead to large systematic errors in \sig, but they should be considered as part of the total \sig\ error budget.

\section{Spearman rank-order correlation results}\label{sec_spearman_rank_order}

In Table \ref{tab_spearman} we present the results of the Spearman rank-order correlation tests we have conducted to look for links between various properties and parameters. 
The table is presented with color codes of $p$-values for visual aid.

\begin{deluxetable*}{|l|l|l|l|l|l|l|l|l|l|l|l|l|l|l|c}
\rotate
\tabletypesize{\footnotesize}
\tablecaption{The Spearman rank-order correlation results. The table is presented with a color code of $\rho$ values for visual aid. Black, blue and red colors correspond to $p$-value$ \leq$ 0.01, $p$-value$\leq$ 0.05, and $p$-value$>$ 0.05, respectively. \label{tab_spearman}}
\tablehead{
\colhead{} & \colhead{\sigs} & \colhead{$(1+z)$} & \colhead{$FWHM_\mathrm{H\beta}$} & \colhead{$L_\mathrm{H\beta}$} & \colhead{$L_\mathrm{5100}$} & \colhead{$M_\mathrm{BH,H\beta}$} & \colhead{$FWHM_\mathrm{H\alpha}$} &      \colhead{\LHa} & \colhead{$M_\mathrm{BH,H\alpha}$} & \colhead{\nh}  & \colhead{\Lbol} & \colhead{\eddr} & \colhead{\AV} &   \colhead{\delm}
}
\startdata
$\sigs$ & \textcolor{red}{1} & \textcolor{black}{0.36} & \textcolor{black}{0.51} & \textcolor{black}{0.27} & \textcolor{black}{0.31} & \textcolor{black}{0.55} & \textcolor{black}{0.48} & \textcolor{black}{0.33} & \textcolor{black}{0.51} & \textcolor{red}{-0.12} & \textcolor{black}{0.47} & \textcolor{black}{-0.21} & \textcolor{red}{0.09} & \textcolor{black}{-0.28} & \\
$\log(1+z)$ & \textcolor{black}{0.36} & \textcolor{red}{1} & \textcolor{black}{0.24} & \textcolor{black}{0.51} & \textcolor{black}{0.54} & \textcolor{black}{0.48} & \textcolor{black}{0.27} & \textcolor{black}{0.66} & \textcolor{black}{0.57} & \textcolor{red}{-0.01} & \textcolor{black}{0.83} & \textcolor{red}{0.06} & \textcolor{red}{0.01} & \textcolor{black}{0.33} & \\
$FWHM_{\mathrm \Hbeta}$ & \textcolor{black}{0.51} & \textcolor{black}{0.24} & \textcolor{red}{1} & \textcolor{red}{-0.01} & \textcolor{red}{0.09} & \textcolor{black}{0.78} & \textcolor{black}{0.68} & \textcolor{red}{0.03} & \textcolor{black}{0.48} & \textcolor{red}{-0.02} & \textcolor{black}{0.24} & \textcolor{black}{-0.43} & \textcolor{black}{0.26} & \textcolor{red}{0.08} & \\
$L_{\Hbeta}$ & \textcolor{black}{0.27} & \textcolor{black}{0.51} & \textcolor{red}{-0.01} & \textcolor{red}{1} & \textcolor{black}{0.88} & \textcolor{black}{0.56} & \textcolor{black}{0.24} & \textcolor{black}{0.85} & \textcolor{black}{0.64} & \textcolor{red}{-0.06} & \textcolor{black}{0.6} & \textcolor{black}{-0.24} & \textcolor{black}{-0.49} & \textcolor{black}{0.42} & \\
$L_{5100}$ & \textcolor{black}{0.31} & \textcolor{black}{0.54} & \textcolor{red}{0.09} & \textcolor{black}{0.88} & \textcolor{red}{1} & \textcolor{black}{0.59} & \textcolor{black}{0.23} & \textcolor{black}{0.78} & \textcolor{black}{0.59} & \textcolor{red}{-0.07} & \textcolor{black}{0.55} & \textcolor{black}{-0.22} & \textcolor{black}{-0.43} & \textcolor{black}{0.33} & \\
$M_{\rm BH,H\beta}$ & \textcolor{black}{0.55} & \textcolor{black}{0.48} & \textcolor{black}{0.78} & \textcolor{black}{0.56} & \textcolor{black}{0.59} & \textcolor{red}{1} & \textcolor{black}{0.68} & \textcolor{black}{0.52} & \textcolor{black}{0.78} & \textcolor{red}{-0.1} & \textcolor{black}{0.53} & \textcolor{black}{-0.49} & \textcolor{red}{-0.09} & \textcolor{black}{0.31} & \\
$FWHM_{\mathrm \Halpha}$ & \textcolor{black}{0.48} & \textcolor{black}{0.27} & \textcolor{black}{0.68} & \textcolor{black}{0.24} & \textcolor{black}{0.23} & \textcolor{black}{0.68} & \textcolor{red}{1} & \textcolor{black}{0.23} & \textcolor{black}{0.79} & \textcolor{red}{-0.04} & \textcolor{black}{0.34} & \textcolor{black}{-0.72} & \textcolor{red}{0.09} & \textcolor{black}{0.39} & \\
$L_{\Halpha}$ & \textcolor{black}{0.33} & \textcolor{black}{0.66} & \textcolor{red}{0.03} & \textcolor{black}{0.85} & \textcolor{black}{0.78} & \textcolor{black}{0.52} & \textcolor{black}{0.23} & \textcolor{red}{1} & \textcolor{black}{0.74} & \textcolor{red}{-0.04} & \textcolor{black}{0.76} & \textcolor{blue}{-0.18} & \textcolor{black}{-0.5} & \textcolor{black}{0.46} & \\
$M_{\rm BH, \Halpha}$ & \textcolor{black}{0.51} & \textcolor{black}{0.57} & \textcolor{black}{0.48} & \textcolor{black}{0.64} & \textcolor{black}{0.59} & \textcolor{black}{0.78} & \textcolor{black}{0.79} & \textcolor{black}{0.74} & \textcolor{red}{1} & \textcolor{red}{-0.05} & \textcolor{black}{0.66} & \textcolor{black}{-0.59} & \textcolor{black}{-0.29} & \textcolor{black}{0.55} & \\
$\nh$ & \textcolor{red}{-0.12} & \textcolor{red}{-0.01} & \textcolor{red}{-0.02} & \textcolor{red}{-0.06} & \textcolor{red}{-0.07} & \textcolor{red}{-0.1} & \textcolor{red}{-0.04} & \textcolor{red}{-0.04} & \textcolor{red}{-0.05} & \textcolor{red}{1} & \textcolor{red}{0.01} & \textcolor{red}{0.09} & \textcolor{red}{0.05} & \textcolor{red}{0} & \\
$\av$ & \textcolor{black}{0.47} & \textcolor{black}{0.83} & \textcolor{black}{0.24} & \textcolor{black}{0.6} & \textcolor{black}{0.55} & \textcolor{black}{0.53} & \textcolor{black}{0.34} & \textcolor{black}{0.76} & \textcolor{black}{0.66} & \textcolor{red}{0.01} & \textcolor{red}{1} & \textcolor{red}{0.09} & \textcolor{red}{0.06} & \textcolor{black}{0.27} & \\
\Lbol & \textcolor{black}{-0.21} & \textcolor{red}{0.06} & \textcolor{black}{-0.43} & \textcolor{black}{-0.24} & \textcolor{black}{-0.22} & \textcolor{black}{-0.49} & \textcolor{black}{-0.72} & \textcolor{blue}{-0.18} & \textcolor{black}{-0.59} & \textcolor{red}{0.09} & \textcolor{red}{0.09} & \textcolor{red}{1} & \textcolor{black}{0.42} & \textcolor{black}{-0.47} & \\
\eddr\ & \textcolor{red}{0.09} & \textcolor{red}{0.01} & \textcolor{black}{0.26} & \textcolor{black}{-0.49} & \textcolor{black}{-0.43} & \textcolor{red}{-0.09} & \textcolor{red}{0.09} & \textcolor{black}{-0.5} & \textcolor{black}{-0.29} & \textcolor{red}{0.05} & \textcolor{red}{0.06} & \textcolor{black}{0.42} & \textcolor{red}{1} & \textcolor{black}{-0.38} & \\
$\delm$ & \textcolor{black}{-0.28} & \textcolor{black}{0.33} & \textcolor{red}{0.08} & \textcolor{black}{0.42} & \textcolor{black}{0.33} & \textcolor{black}{0.31} & \textcolor{black}{0.39} & \textcolor{black}{0.46} & \textcolor{black}{0.55} & \textcolor{red}{0} & \textcolor{black}{0.27} & \textcolor{black}{-0.47} & \textcolor{black}{-0.38} & \textcolor{red}{1} & \\
\enddata
\end{deluxetable*}

\clearpage
\newpage

\section{Principal Component Analysis}\label{sec:PCA}

To identify the main parameters that are driving the variance in our data set, we conduct a Principal Component Analysis (PCA) focusing on the observable parameters as follows: \sig\/, $\log(1+z)$, $FWHM_\mathrm{H\alpha}$, $\log{L_\mathrm{H\alpha}}$, \av\/, $L_\mathrm{Bol}$. Interestingly, the first three eigenvectors (EV1 and EV2) explain 83.3\% variance in the data set. The most dominant one is the first eigenvector, which explains 48.6\% variance. The remaining three eigenvectors explain 8.5\%, 6.1\%, and 2.1\% variance, respectively. By inspecting the correlations of these eigenvectors with the selected variables (see Table~ \ref{tab:PCA}), we find that the first and second eigenvectors are mostly driven by the anti-correlation between $\av$ and $L_\mathrm{H\alpha}$ indicating an obscuration effect in the observed $L_\mathrm{H\alpha}$, in agreement with previous works \citep[C20;][]{Mejia_Broadlines,Ricci_DR2_NIR_Mbh}. The correlations between $z$, luminosity ($L_\mathrm{Bol}$), and \sig\ mostly drive the third eigenvector. This traces the selection bias resulting from the flux-limited nature of the sample, which favors the detection of high luminosity and high \sig\ objects at higher redshifts. The remaining eigenvectors only explain the 16.7\% variance of the data set, and there are at least four dominant parameters, which lay the responsibility for each eigenvector. This indicates a significant scatter in the measured properties suggesting that different parameters than the first, second, and third eigenvectors are the main drivers for the remaining eigenvectors.

\begin{table*}[hb!]
\centering
\caption{The resulting Spearman correlation coefficients between the select features for each Eigenvector (EV) from Principal Component Analysis.}
\begin{tabular}{cccccccc}
\hline
\hline
 Feature & EV 1 & EV 2 & EV 3 & EV 4 & EV 5 & EV 6   \\
 Variance & 48.6\% & 20.9\% & 13.8\% & 8.5\% & 6.1\% & 2.1\% \\
 \hline
$\log   \sig$                      & 0.22  & -0.47 & 0.82 & -0.19 & 0.19  & -0.26 \\
$\log(1+z)$                        & -0.29 & -0.02 & 0.79 & 0.03  & 0.09  & -0.60 \\
$\log FWHM_\mathrm{H\alpha}$       & 0.39  & -0.63 & 0.08 & -0.17 & 0.74  & -0.23 \\
$\log L_\mathrm{H\alpha}$          & -0.79 & 0.66  & 0.53 & 0.06  & 0.31  & -0.37 \\
\av\ & 0.49  & -0.81 & 0.21 & -0.27 & -0.39 & -0.39 \\
$\log L_\mathrm{Bol}$              & -0.43 & 0.15  & 0.74 & 0.03  & 0.36  & -0.54 \\
\hline
\end{tabular}
\label{tab:PCA}
\end{table*}

\end{appendix}

\global\suppressAffiliationsfalse
\allauthors

\end{document}